\documentclass[sigconf, nonacm]{acmart}

\AtBeginDocument{%
  }

\usepackage{hyperref}
\usepackage[shortlabels]{enumitem}
\usepackage{amsmath}
\usepackage{amsfonts}
\usepackage{graphicx}
\usepackage{textcomp}
\usepackage{xcolor}
\usepackage{multirow}
\usepackage{balance}
\makeatletter
\newcommand*{\rom}[1]{\expandafter\@slowromancap\romannumeral #1@}
\makeatother
\usepackage{booktabs}
\usepackage{balance}
\usepackage{array}
\usepackage{multirow}
\usepackage{tabularray}
\usepackage{tcolorbox}
\usepackage{color,soul}
\usepackage[normalem]{ulem}
\useunder{\uline}{\ul}{}

\usepackage{appendix}
\usepackage{etoolbox}
\BeforeBeginEnvironment{appendices}{\clearpage}

\setlist[enumerate]{leftmargin=*}
\setlist[itemize]{leftmargin=*}

\usepackage{caption}
\usepackage{xcolor}

\newcommand{\rv}[1]{{\color{black}{#1}}}

\begin{document}

\title{SmartLLMs Scheduler: A Framework for Cost-Effective LLMs Utilization}

\author{Yueyue Liu}
\email{yueyue.liu@uon.edu.au}
\affiliation{%
  \institution{University of Newcastle}
  \streetaddress{Callaghan}
  \city{Newcastle}
  \state{NSW}
  \country{Australia}
  \postcode{2308}
}

\author{Hongyu Zhang}
\authornote{Corresponding author.}
\email{hyzhang@cqu.edu.cn}
\affiliation{%
  \institution{
  Chongqing University}
  \streetaddress{Chongqing}
  \city{Chongqing}
  \country{China}
  \postcode{401331}
}

\author{Yuantian Miao}
\email{sky.miao@newcastle.edu.au}
\affiliation{%
  \institution{University of Newcastle}
  \streetaddress{Callaghan}
  \city{Newcastle}
  \state{NSW}
  \country{Australia}
  \postcode{2308}
}

\begin{abstract}
Large Language Models (LLMs) such as GPT-4 and Llama have shown remarkable capabilities in a variety of software engineering tasks. Despite the advancements, their practical deployment faces challenges, including high financial costs, long response time, and varying performance, especially when handling a large number of queries (jobs).
Existing optimization strategies for deploying LLMs for diverse tasks focus on static scheduling, which requires extensive training data for performance prediction, increasing the computational costs and limiting the applicability and flexibility.
In this paper, we propose the SmartLLMs Scheduler (SLS), a dynamic and cost-effective scheduling solution. The key idea is to learn LLMs' performance on diverse tasks and incorporate their real-time feedback to update strategies periodically. Specifically, SLS incorporates three key components, including an Adaptive Cache Manager, a Performance-Cost Optimized Scheduler, and a Dynamic Update Manager. The Cache Manager stores the outputs of previously processed queries and employs an adaptive strategy to reduce redundant computations and minimize response times. For queries not found in the cache, the Scheduler dynamically allocates them to the most suitable LLM based on the predicted performance and cost from models that take both query-specific and LLM-specific features as input. The Update Manager continuously refines the cache and scheduling strategies based on real-time feedback from the assigned queries
to enhance decision-making and adapt to evolving task characteristics.
To evaluate the effectiveness of SLS, we conduct extensive experiments on two LLM-based software engineering tasks, including log parsing and code generation. The results show that SLS significantly outperforms the baseline methods, achieving an average performance improvement of 198.82\% and an average processing time reduction of 63.28\%. It also yields an average cost reduction of 69.70\% in scenarios where savings are achieved.

\end{abstract}


\maketitle

\section{Introduction}
\label{sec1_introduction}

Large Language Models (LLMs) have garnered significant attention from the software engineering community due to their remarkable language understanding and generation capabilities~\cite{zheng2024harnessing, baek2024knowledge}. A growing number of LLM services with varying levels of performance and pricing are accessible, such as commercial LLMs GPT-4 from OpenAI\footnote{\label{openai}https://openai.com} and Claude from Anthropic\footnote{\label{anthropic}https://www.anthropic.com}. Additionally, open-source alternatives, such as LLaMA3 from Meta\footnote{\label{meta}https://ai.meta.com}, provide more flexible deployment options. The versatility of LLMs has enabled a wide range of applications in intelligent software engineering, such as code generation~\cite{fakhoury2024llm, wu2024unify, nam2024using}, fault detection~\cite{xiao2024free, xu2024divlog, roy2024exploring}, and bug fixing~\cite{fu2024missconf, kang2023large, xia2024automated}.

Despite the above-mentioned advancement, the practical utilization of LLM for handling large volumes of queries across various tasks remains challenging. 
These challenges arise from high financial costs, long response times, inconsistent performance across different tasks, and the lack of balance among these factors~\cite{hou2023large, chen2024frugalgpt}. 
First, using LLMs can be expensive, particularly for tasks requiring a large number of queries~\cite{chen2024frugalgpt,vsakota2024fly}. For instance, utilizing Claude 3 Opus, a commercial model developed by Anthropic, to process 100,000 customer support queries with 500 tokens per query would cost approximately \$1,687 based on the current pricing model~\cite{anthropic_pricing}.
Beyond financial costs, LLM API response times can also be substantial, especially during peak usage periods or when handling a high volume of queries~\cite{wang2024survey}. High latency poses challenges for real-time applications, especially time-sensitive tasks.
Another challenge is the inconsistent performance of LLMs across different tasks. While large models like GPT-4 are expected to have higher computational capacity and achieve better performance, they do not always outperform smaller models, particularly in specialized or domain-specific tasks~\cite{zhong2024logparser}. Conversely, for less complex tasks, smaller LLMs can achieve performance comparable at a lower computational cost~\cite{chen2024frugalgpt}, whilst being limited in handling tasks' high complexity and variety.

Given these challenges, scheduling methods have emerged as a promising approach for optimizing the utilization of LLMs. It allocates queries across multiple models to balance optimization objectives such as cost, response time, and accuracy~\cite{ chen2024frugalgpt}.  
{Current strategies predominantly utilize static scheduling, where the allocation of all queries is determined before submission based on initial predictions without incorporating real-time feedback from query execution~\cite{liu2024optllm, vsakota2024fly}.} For example, FORC~\cite{vsakota2024fly} predicts the cost and the probability of a query being successfully processed by each candidate LLM and selects the most suitable one based on a cost-performance tradeoff. Similarly, OptLLM~\cite{liu2024optllm} adopts a multi-objective optimization approach, integrating a multi-label classification model with uncertainty estimation. It generates a set of optimal solutions from which users can select based on their preferred tradeoff between expected performance and cost.

\begin{figure}
    \centering
    \includegraphics[width=1\linewidth]{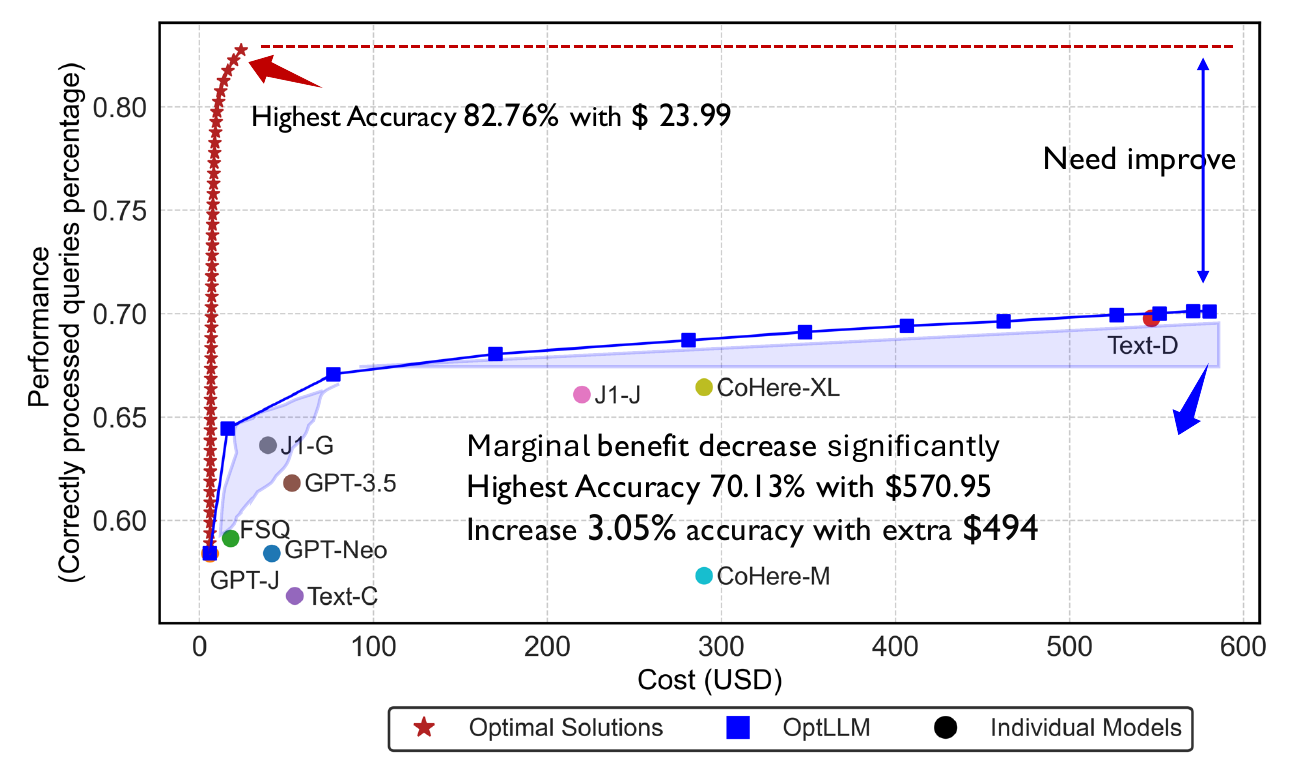}
    \caption{Limitations of Static Scheduling (OptLLM) in LLM Utilization Optimization}
    \label{limitation}
    \vspace{-18pt}
\end{figure}

However, static scheduling strategies, such as those employed by FORC~\cite{vsakota2024fly} and OptLLM~\cite{liu2024optllm}, inherently limit flexibility and adaptability, as decisions remain fixed once queries are allocated. Such strategies are heavily dependent on the accuracy of the prediction model, and inaccurate predictions may lead to suboptimal query allocations since adjustments are not considered at runtime. For example, OptLLM, despite adopting multi-objective optimization and uncertainty estimation, struggles to significantly enhance accuracy beyond a certain performance threshold, with incremental cost increases yielding only marginal improvements, reflecting its inherent limitations, which is illustrated in Figure~\ref{limitation}.
Achieving reliable predictions typically requires extensive training data. For instance, FORC~\cite{vsakota2024fly} utilizes extensive query data (five times more than the evaluation set) for its prediction model. This data collection is costly, as queries must be submitted to all candidate LLMs to provide `sufficient' information before scheduling. Beyond these limitations, existing methods also fail to incorporate result reuse for repeated/similar queries, leading to redundant computations, especially in tasks with a high number of repeated/similar queries. 

To address these limitations, this paper presents a framework called SmartLLMs Scheduler (SLS) for dynamic and cost-effective query scheduling on LLMs to save costs, reduce makespan (i.e., total time to finish all queries), and enhance LLM utilization with a real-time feedback mechanism. Unlike static scheduling, which assigns pre-collected queries at once, SLS employs a rolling horizon model that dynamically schedules incoming queries at fixed intervals. 
It continuously learns LLM performance across different task types and periodically updates scheduling strategies with feedback from executed queries.
To achieve this, SLS includes three key components: an Adaptive Cache Manager, a Performance-Cost Optimized Scheduler, and a Dynamic Update Manager. The Cache Manager retrieves previously processed queries to minimize redundant computations, reducing computational costs for repeated or similar queries. The Scheduler dynamically allocates queries based on predicted performance and cost, incorporating customizable rules to balance accuracy and efficiency. The Dynamic Update Manager continuously monitors the prediction model and caching performance at each period and updates them with feedback from submitted queries when prediction accuracy or caching hits a predefined threshold.
By using real-time feedback from executed queries, SLS reduces reliance on the initially trained prediction model, ensuring continuous learning and adaptability to real-world workloads.

To evaluate the effectiveness of SLS, we conduct extensive experiments on two LLM-based software engineering tasks, namely log parsing~\cite{ma2024llmparser, xu2024divlog} and code generation~\cite{jiang2024survey, mu2024clarifygpt}. Specifically, we use three datasets for log parsing~\cite{zhu2023loghub} and one for code generation~\cite{yin2018learning}. Compared to baselines, SLS achieves significantly higher accuracy. \rv{For log parsing, accuracy improves by 0.07\% to 2894.05\%, with cost savings ranging from 5.99\% to 98.39\% in scenarios where savings are achieved. Similarly, for code generation, accuracy improves by 6.72\% to 2838.39\%, while cost savings range from 32.24\% to 91.31\% in cost-reducing cases.}
To assess the generalizability, we further evaluate SLS on two Natural Language Processing (NLP) datasets for text classification and obtain improved results. 

Our major contributions are as follows:
\begin{enumerate}
    \item We present SLS, a novel dynamic scheduling framework designed to optimize the utilization of LLMs. SLS leverages adaptive caching to store and reuse previous results, dynamically allocates LLMs to incoming queries based on predicted performance and cost, and periodically updates its strategies with real-time feedback at the end of each period to improve future allocations.
    \item We conduct a comprehensive evaluation of SLS on two LLM-based software engineering tasks (log parsing and code generation). The results show that SLS consistently outperforms baseline methods in performance, cost, and makespan, demonstrating its effectiveness and flexibility for real-world LLM utilization. Additionally, an experiment on text classification further illustrates the generalizability of SLS.
\end{enumerate}

\section{Related Work}
\label{sec2_background}

Researchers have explored various techniques to optimize the performance and efficiency of LLMs while reducing the financial and computational costs of their deployment~\cite{yadav2024pag, xia2024llm, sun2024consistency}.  
One common approach is caching, where previously processed queries and their corresponding responses are stored for reuse, reducing the inference time and invocation cost on LLMs service~\cite{bang2023gptcache}. For example, GPTCache~\cite{bang2023gptcache}, a conventional caching framework, stores previously computed results and attempts to match future queries exactly based on simple heuristics. While this approach can save response times and API costs, it often struggles to recognize semantically similar queries, leading to a high rate of false misses. Some caching frameworks address additional constraints, such as privacy protection and memory efficiency. For example, MeanCache\cite{gill2024privacy} leverages federated learning to protect user data, focusing on privacy concerns. The framework proposed by Zhang et al.~\cite{zhangcam} emphasizes memory efficiency by compressing embeddings to reduce storage requirements. These specialized caching solutions enhance cost and performance while addressing domain-specific challenges.

Another important technique is scheduling, which involves allocating LLM to queries based on task complexity and constraints such as budget or response time. Several frameworks have been proposed for the LLM invocation optimization problem~\cite{chen2024frugalgpt,vsakota2024fly}. For example, Chen et al.\cite{chen2024frugalgpt} introduced FrugalGPT, a framework that selects LLMs for incoming queries by sequentially invoking models until a predefined performance threshold is reached. This approach reduces costs while maintaining accuracy. However, a potential drawback is that multiple invocations for a single query can increase costs, particularly if cheaper models perform poorly. More broadly, these frameworks often depend on performance prediction models that require substantial amounts of labeled data for training, leading to high annotation costs~\cite{vsakota2024fly}. \rv{To address this limitation, Shnitzer et al.~\cite{shnitzer2023large} developed a framework that leverages existing benchmark datasets to train correctness predictors for routing queries to the most suitable LLM. Although this approach does not involve conventional transfer learning, it effectively repurposes prior evaluation data to make informed predictions on new tasks, thereby reducing runtime costs without requiring per-query LLM evaluations. Similarly, Smoothie~\cite{guha2024smoothie} achieves label-free routing by prompting a lightweight LLM to both generate an answer and assess its own response. This self-assessed confidence score is then used to determine whether the query should be escalated to a more powerful model, enabling efficient and adaptive routing without the need for ground-truth labels.}

In addition to these single-objective approaches, Liu et al.~\cite{liu2024optllm} proposed OptLLM, which formulates the query allocation problem for LLMs as a multi-objective optimization task and provides a set of solutions to balance cost and performance. To mitigate the impact of inaccurate predictions misguiding the search algorithm, OptLLM employs a multi-label classification model with uncertainty estimation to predict the performance of candidate LLMs for each query. However, OptLLM focuses on static scheduling, which limits its adaptability to dynamic workloads. Moreover, it does not fully leverage information from previously processed queries, which could further improve performance.

Our proposed SLS differs from OptLLM in several key ways. First, while OptLLM provides multiple optimal solutions for users to choose from, SLS focuses on adaptive and dynamic scheduling, enabling real-time decision-making as queries arrive. This makes SLS more suitable for scenarios where jobs arrive continuously or unpredictably. Second, SLS incorporates an adaptive caching mechanism, which is absent in OptLLM, to reduce redundant computations by reusing previously computed results. This significantly reduces both processing time and overall system cost. Lastly, instead of multi-objective optimization, SLS applies a performance-cost ratio to make it adaptable to a dynamic framework. It informed by query similarities and predictions from separate performance and cost models, aim to reduce makespan via cache reuse and balance cost and performance by selecting the LLM that offers the highest predicted performance-to-cost ratio for each query.

\section{Problem Statement}
\label{sec3_problem_statement}
\subsection{Problem Overview}
Assume that a set of LLMs with different capabilities and costs are accessible via APIs, while queries arrive dynamically over time to be processed. SLS aims to allocate each query to the most suitable LLM to minimize overall cost while maximizing accuracy.

\noindent\textbf{Job Requests.}  
This paper adopts a rolling horizon scheduling approach~\cite{glomb2022rolling}, where the scheduling process runs in discrete horizons \( \{T_1, T_2, \dots, T_T \}\), each covering a fixed time interval (e.g., 1 hour). Let \( J_t \) be the set of jobs that arrive in the scheduling period \( t \). Each query \( j_i^t \) arriving in \( T_t \) is characterized by its token count \( tn_i^t \).

\noindent\textbf{Candidate LLMs.}  
The user has access to a set of candidate LLMs:
\begin{equation}
L=\{l_1,l_2,\dots,l_m\}
\end{equation}
Each LLM \( l_k \) has an input token price \( price_k^{\text{input}} \) and a generation token price \( price_k^{\text{output}} \).  
Let \( tn_i \) denote the number of input tokens in query \( j_i \), and \( tn_o \) represent the number of generated tokens in the response. The cost of processing query \( j_i \) on LLM \( l_k \) is determined by both the input and generated token counts, computed as:
\begin{equation}
cost_{i,k} = tn_i \times price_k^{\text{input}} + {tn}_o \times price_k^{\text{output}}
\end{equation}

\subsection{Per-Horizon Optimization Problem}
For each scheduling horizon \( t \), queries must be assigned to LLMs to optimize cost and accuracy.  

\noindent{Decision Variable:}  
\begin{equation}
x_{i,k}^t =
\begin{cases} 
1, & \text{if query } j_i^t \text{ is assigned to LLM } l_k \\
0, & \text{otherwise}
\end{cases}
\end{equation}

\noindent{Objective Functions:}
\begin{equation}
\text{Minimize Cost:} \quad f_{\text{cost}}^t = \sum_{i \in J_t} cost_i^t
\end{equation}
\begin{equation}
\text{Maximize Performance:} \quad f_{\text{perf}}^t = \frac{1}{n_t} \sum_{i \in J_t} perf_i^t
\end{equation}
Subject to:
\begin{equation}
cost_i^t = \sum_{k=1}^{m} x_{i,k}^t \times cost_{i,k}^t, \quad \forall i \in J_t
\end{equation}
\begin{equation}
perf_i^t = \sum_{k=1}^{m} x_{i,k}^t \times perf_{i,k}^t, \quad \forall i \in J_t
\end{equation}
\begin{equation}
\sum_{k=1}^{m} x_{i,k}^t = 1, \quad \forall i \in J_t
\end{equation}
\begin{equation}
x_{i,k}^t \in \{0,1\}, \quad \forall i \in J_t, \forall k \in L
\end{equation}

While optimization is performed at each scheduling horizon, we assess the overall effectiveness of SLS using global evaluation metrics, including total cost and overall performance across all scheduling intervals. These evaluation metrics are further detailed in Section~\ref{sec5Evaluation metric}.

\section{Methodology}
\label{sec3:methodology}
\begin{figure*}
    \centering
    \includegraphics[width=0.9\linewidth]{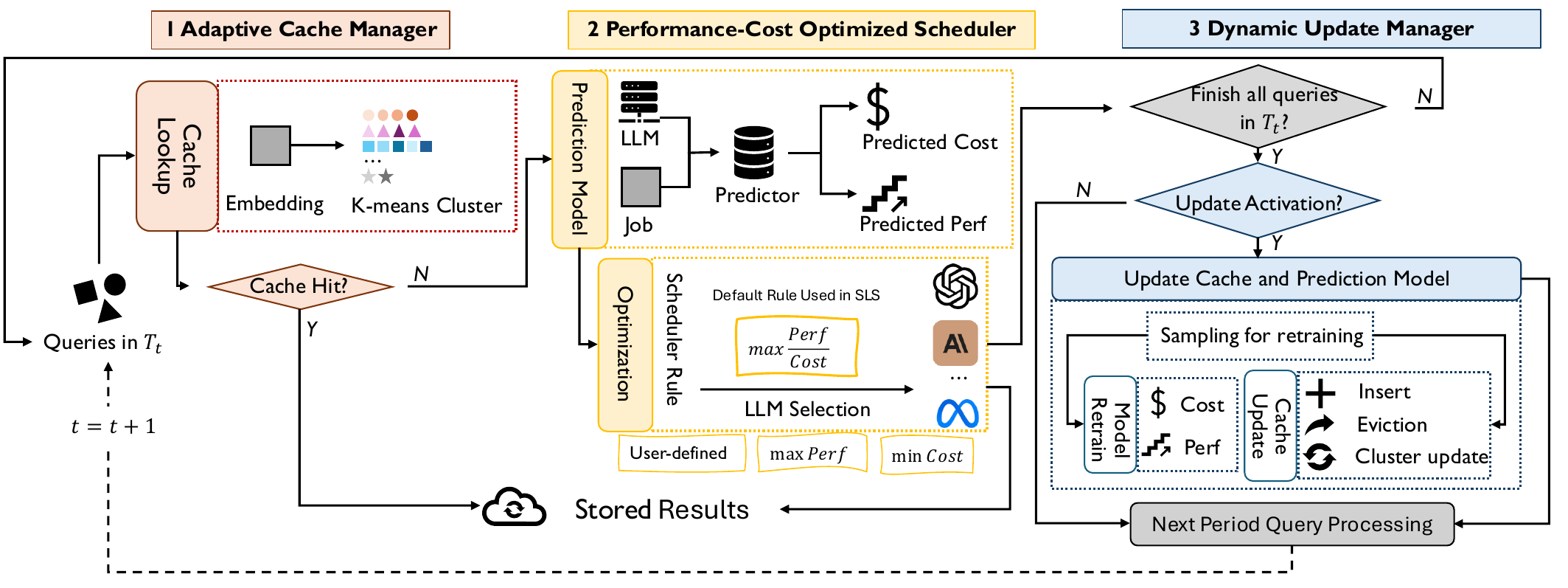}
    \vspace{-6pt}
    \caption{ The Overall Workflow of SLS}
    \vspace{-6pt}
    \label{fig:slmframework}
\end{figure*}

\subsection{Overview}
SLS is a framework designed to efficiently utilize LLMs. It consists of an Adaptive Cache Manager, a Performance-Cost Optimized Scheduler, and a Dynamic Update Manager. Specifically, the Adaptive Cache Manager reuses stored results to minimize redundant computations, while the Scheduler adaptively selects the most suitable LLMs based on predicted performance and cost. The Dynamic Update Manager periodically adjusts model parameters and retrains the prediction model based on real-time feedback from the results of invoked LLM for submitted queries. The overall workflow is shown in Figure~\ref{fig:slmframework}.

\subsection{Adaptive Cache Manager} 
\label{sec3:cache}
The Adaptive Cache Manager in SLS minimizes redundant LLM invocations by utilizing query embeddings to reuse previously processed results. By representing each query as an embedding vector and grouping them into clusters, SLS captures contextual information and effectively identifies similarities between new and stored queries. The following subsections detail the key components of the Cache Manager.

\subsubsection{Query embedding representation}
The Cache Manager leverages query embeddings to convert inputs into high-dimensional numerical representations for similarity comparisons. To ensure accurate representation and similarity detection, SLS classifies incoming queries as text or code and applies separate pre-trained language models optimized for each type.

For text-based queries, SLS utilizes the SentenceTransformer model~\cite{Reimers2019SentenceBERTSE}, which is optimized for capturing nuanced semantic relationships between sentences, making it effective for NLP tasks such as text classification~\cite{su2023multi}. 
For code-based queries, SLS adopts CodeBERT~\cite{feng-etal-2020-codebert}, which is pre-trained on a diverse corpus of programming languages. CodeBERT can capture the syntactic and structural nuances of code, making it particularly suitable for code-related tasks such as code generation and code summarization~\cite{mashhadi2021applying, karmakar2021pre}. 

These models enable SLS to generate embeddings that capture the key characteristics and contextual information of each query, allowing the cache to identify relevant matches based on the underlying meaning, rather than relying solely on surface-level textual or structural similarities. While these models are used as defaults, SLS also allows users to integrate alternative models that better align with specific application requirements or domain-specific needs.

\subsubsection{Similarity calculation}
The Cache Manager employs Cosine similarity to measure the content similarity between queries based on their embedding representations. Cosine similarity ensures that differences in scale, such as query length, do not impact the similarity score. This makes it effective for comparing queries with similar meanings or contexts, even if they differ in length or structure. The similarity score is computed as follows:
\[
\text{Cosine Similarity} = \frac{\mathbf{A} \cdot \mathbf{B}}{\|\mathbf{A}\| \|\mathbf{B}\|}
\]
where $\mathbf{A}$ and $\mathbf{B}$ are the embedding vectors of the queries. A higher score indicates a stronger similarity between the two queries.

\subsubsection{Embedding clustering}
\label{sec3:embedding cluster}
To enhance the cache matching efficiency, SLS utilizes K-Means clustering to organize query embeddings into distinct groups. When the number of stored embeddings reaches a predefined threshold, the clustering process is triggered, and the query embeddings are partitioned into clusters based on their spatial distribution in the embedding space. Instead of searching the entire cache, this strategy enables SLS to focus similarity comparisons within a smaller subset of related queries, improving retrieval speed and accuracy.

\subsubsection{Adaptive similarity threshold adjustment}
\label{sec3:threshold in cache}
The Cache Manager adaptively adjusts the similarity threshold $\tau$ based on two performance metrics: the cache success rate, which measures the proportion of correct results returned, and the cache hit rate, which indicates how often a match is retrieved regardless of correctness. Based on these metrics, the similarity threshold is modified through two key strategies: 
\begin{itemize}
    \item Threshold tightening: The similarity threshold is increased (tightened) when the cache success rate is low, indicating that the cache does not frequently return correct results. This situation may arise when the cache retrieves many incorrect entries, despite a high cache hit rate. Tightening the threshold makes the cache more selective, reducing false positives and improving precision by focusing only on highly similar entries.
    \item Threshold relaxation: The similarity threshold is decreased (relaxed) when the cache success rate is consistently high, suggesting that the cache effectively retrieves correct results. Lowering the threshold broadens the search space, allowing the cache to include entries with slightly lower similarity scores. This strategy increases cache coverage, enhancing the likelihood of retrieving relevant matches for new queries and improving overall recall.
\end{itemize}

The cache hit rate is directly measurable, while the cache success rate is estimated based on a small number of inspected jobs when the cache update mechanism is triggered. The Cache Manager maintains an effective balance between precision and recall by dynamically adjusting the similarity threshold based on these performance indicators. 

\subsubsection{Cache maintenance and expansion}
\label{Cache maintenance and expansion}
The cache performs three primary operations: cache insertion, cluster update, and cache eviction, to optimize query retrieval efficiency and maintain the relevance of stored entries under constrained cache capacity.
\begin{itemize}
\item Cache insertion: Each new query, along with its result and embedding, is stored in the cache using a unique identifier. The access frequency and timestamp are also recorded to assist with future replacement decisions.
\item Cluster update: The cache applies K-means clustering to group stored embeddings, improving similarity-based lookups and reducing retrieval complexity as new entries are added.
\item Cache eviction: When the cache reaches capacity, the least recently used (LRU) entry is removed, ensuring that only the most relevant and frequently accessed queries are retained.
\end{itemize}

\subsection{Performance-Cost Optimized Scheduler}
The Performance-Cost Optimized Scheduler allocates queries that result in a cache miss, directing them to the most suitable LLM based on predefined scheduling rules, and satisfying user preferences for performance and cost.
It comprises two main components: the Prediction component and the Optimization component. The Prediction component uses machine learning models to estimate the expected cost and the probability of successful completion for each query across different candidate LLMs. The Optimization component leverages these predictions to select proper LLM according to a scheduling rule, such as maximizing the performance-to-cost ratio, ensuring that queries are processed cost-effectively while maintaining high-quality outcomes.

\subsubsection{Prediction component}
The prediction model is essential due to the inherent uncertainty in determining whether an LLM can successfully process a given query before it has been executed. This uncertainty arises from the variability in the complexity and requirements of individual queries, which can differ performance outcomes across LLMs. Moreover, while the input cost can be estimated from the number of tokens and the model unit price, the output cost, which depends on the number of tokens required to generate a complete response, remains unknown until the query is processed. 
Based on historical performance data, the prediction model can estimate the probability of successful query completion and the associated costs, providing essential information for selecting the most suitable LLM.

\textbf{Feature extraction.} Current prediction models for LLM invocation optimization only considered query features~\cite{vsakota2024fly, liu2024optllm}, neglecting the influence of the LLM itself on performance and cost. While query characteristics significantly affect outcomes, different LLMs exhibit varying performance and cost. To address this limitation, SLS incorporates features from both the query and the LLM.

SLS uses the embedding vector generated by the Cache Manager to capture the semantic and contextual representation of the query. For LLM-specific characteristics, a one-hot encoding represents the selected model, allowing the scheduler to factor in variations in cost and performance across LLMs. The combined query and LLM features are normalized with a unified scaler to ensure consistency and comparability in the final representation.

\textbf{Prediction models.} Then two separate prediction models are constructed: one for cost estimation and another for performance prediction. This study utilizes XGBoost models for both tasks. Specifically, an XGBoost classifier is used to predict the probability of successful query resolution by the LLM, utilizing a binary logistic objective to distinguish between successful and failed queries. Simultaneously, an XGBoost regressor estimates the associated cost, employing a squared error objective to minimize the difference between the predicted and actual costs. 

Being a flexible framework, SLS allows users to define their own classifiers \rv{(e.g., to predict whether a given LLM can successfully solve a query)} or regression models \rv{(e.g., to estimate performance metrics like BLEU, or to predict cost)}, enabling customization based on specific application needs or performance requirements.

\subsubsection{Optimization component} Then, the optimization component allocates queries efficiently across available LLMs based on the predicted cost and performance, achieving a balance between performance and cost. It defines the objective function and leverages predefined scheduling rules to guide the selection of the most suitable LLM for each job.

\textbf{Optimization objective functions.}
Assume there are $m$ candidate LLMs, and $n$ queries arriving dynamically within the current period $T_t$. The goal is to minimize the total cost and maximize the overall performance. 

Recall the optimization objectives in Section~\ref{sec3_problem_statement}, the total cost includes both the input (prompt) and output (generation) costs for each query assigned to an LLM. Assume $token_i$ is the input token count of query $i$, $price_j^I$ is the input price per token for LLM $j$, and $output_{i,j}$ is the output token count generated by LLM $j$. However, the number of generated tokens \( tn_o \) is unknown before execution, a predicted value \( \hat{tn}_o \) is used during scheduling to estimate the cost.

\begin{equation}
cost_{i,k} = tn_i \times price_k^{\text{input}} + \rv{\hat{tn}_o}\times price_k^{\text{output}}
\end{equation}

The performance objective is to maximize the average success rate of the jobs processed. The performance \(perf_{i,j}\) represents the predicted success probability that query \(i\) will be successfully processed by LLM \(j\). For each job \(i\), the performance is given by:
\begin{equation}
perf_{i} = \sum^{m}_{k=1} x_{i,k} \cdot perf_{i,k}, \quad \forall i \in \{1, \dots, n\}
\end{equation}
where \(x_{i,k} \in \{0, 1\}\) indicates whether query \(i\) is assigned to LLM \(k\).

The average performance across all queries is then defined as:
\begin{equation}
f_{perf} = \frac{1}{n} \sum_{i=1}^{n} perf_{i}
\end{equation}

In addition, each job \(i\) must be assigned to exactly one LLM, ensuring that no job is left unassigned or allocated to multiple models within the same period:
\begin{equation}\label{constraint_assignment}
\sum_{k=1}^{m} x_{i,k} = 1, \quad \forall i \in \{1, \dots, n\}
\end{equation}

\textbf{Scheduling rules for LLM selection}
The scheduling rules define the optimization criteria for LLM selection for each query. These rules reflect different preferences for performance and cost, allowing SLS to align with the specific requirements or priorities. 

The default scheduling rule maximizes the performance-to-cost ratio to ensure the system achieves the maximum value per cost unit, promoting cost-effective query handling. This ratio reflects the trade-off between predicted performance and cost, helping select the most suitable LLM for each query. The performance-to-cost ratio for a given query \(i\) on LLM \(k\) is calculated using the predicted performance \(perf_{i,k}\) and the associated cost \(cost_{i,k}\) as defined by the following equation:

\begin{equation}\label{ratio}
\text{Performance-to-Cost Ratio}_{i,k} = \frac{perf_{i,k}}{cost_{i,k}}
\end{equation}

In addition to this ratio-based rule, SLS offers the Performance-First (Max-Perf) and Cost-Optimized (Min-Cost) strategies, detailed in Section~\ref{sec6discussion}. The framework also supports customizable rules, allowing users to introduce new criteria to adapt to diverse requirements and workloads.

\subsection{Dynamic Update Manager} The Dynamic Update Manager orchestrates periodic updates to accommodate the continuous arrival of new queries and ensure that SLS remains adaptable to the query characteristics. It manages the timing and activation of updates within the framework, covering both the cache and prediction models.

\subsubsection{Adaptive process activation}
The adaptive process activation mechanism governs the timing and necessity of updates within both the Cache Manager and the Scheduler’s prediction models.  Rather than engaging in continuous \rv{inspection and monitoring, which may cause substantial computational overhead and require human-involved effort,} the system adopts a controlled update strategy to balance responsiveness with operational efficiency.

After all jobs in the current period have been completed, the Update Manager assesses whether the system should enter the update process. Specifically, the adaptive process is triggered when the following condition is satisfied:
\begin{equation}\label{mod}
P \mod Q = 0
\end{equation}
where $P$ is the current processing period number and $Q$ controls the update frequency. A larger $Q$ reduces the frequency of updates, helping to prevent excessive recalculations, while a smaller $Q$ increases the frequency, ensuring more frequent updates.

\rv{If the condition is satisfied, the Update Manager performs a lightweight inspection by selecting a subset of the jobs submitted to the LLMs and comparing their outputs against the ground truth.} If a job is correctly handled by the selected LLM, its query and output are considered for inclusion in the cache following the cache update strategy. Regardless of correctness, the inspected samples are retained as candidates for updating the prediction models. The details of how data is selected for model updates are introduced in Section~\ref{subsection:data sampling}.

\rv{Upon completion of the inspection phase, the system determines whether further adaptive adjustments are necessary.} For the cache, the Cache Manager considers adjusting the similarity threshold based on the evaluation criteria described in Section~\ref{sec3:threshold in cache}.
For the prediction models in the Scheduler, the system maintains two separate models: one for estimating success probability and another for the invocation cost. Their performance is evaluated using prediction accuracy and mean absolute error (MAE), respectively.
If the prediction error exceeds a predefined threshold or significant deviations are detected, the system retrains the corresponding models to ensure prediction accuracy and maintain system robustness.

\subsubsection{Data sampling for updates}
\label{subsection:data sampling}
\rv{Once the inspection process is completed, the results are used to update both the cache and the prediction models. To ensure a representative and diverse sample, the selection is guided by two main criteria: the distribution of submitted LLMs and the variation among queries.

The initial training data includes each query along with responses from all candidate LLMs, offering a balanced feature distribution. However, jobs processed during runtime typically only include the response from the selected LLM. To mitigate imbalance during sampling, we select an equal number of jobs from each LLM. Furthermore, query diversity is preserved by measuring embedding distances between queries, ensuring that the sampled data represents a wide range of query characteristics and reducing the risk of overfitting to similar inputs.}

For queries successfully answered by the LLM, the corresponding correct response is added to the cache based on the defined maintenance policy, which incorporates LRU-based eviction and cluster-based query organization using K-Means clustering (see Section~\ref{Cache maintenance and expansion}). The performance prediction model is updated using both successful and failed queries.

\section{Experimental Design}
\label{sec5:Experimental Design}
\subsection{Research questions}
We aim to answer the following research questions (RQs):\\ 
\textbf{RQ1}: How effective is SLS in scheduling LLM jobs? \\ 
\textbf{RQ2}: How do the major components of SLS contribute to its overall performance? \\
\textbf{RQ3}: What factors influence the generalizability of SLS's performance across different applications? \\
\textbf{RQ4}: How do different parameter settings affect SLS's performance?

\subsection{Evaluation tasks and benchmark overview}
To demonstrate the effectiveness of SLS, we conduct extensive experiments on two typical software engineering tasks.

\subsubsection{Log parsing}
Log parsing is to extract structured information from the unstructured log data generated by software systems. Log messages are lines of text that record system events. These messages often include dynamic variables that vary across instances of the same event type, making direct analysis challenging. Log parsing addresses this issue by normalizing the dynamic variables and replacing them with placeholders, creating static templates representing the common structure and content of log messages for a given event type. 
For example, consider the following log message:  
\texttt{2023-10-13 12:34:56 ERROR: Process f3e2 write to /etc/smartd.conf failed.} This log message can be transformed into the following template:  
\texttt{\textless{}timestamp\textgreater{} \textless{}log\_level\textgreater{}: Process \textless{}process\_id\textgreater{} write to \textless{}file\_path\textgreater{} failed.} Recently, some works have proposed LLM-based log parsers and observed better parsing accuracy ~\cite{ma2024llmparser, xu2024divlog}.

We utilize the LogHub-2.0 benchmark dataset~\cite{zhu2023loghub}, which includes logs from 14 diverse open-source systems, such as operating systems, standalone software, and server-side applications. Compared to its predecessor, LogHub-1.0, this version expands the dataset size to over 50 million logs and introduces 3,488 distinct log templates. We selected three systems from the LogHub-2.0 dataset: Apache (Server Application, 51,977 logs), Linux (Operating System, 23,921 logs), and Proxifier (Standalone Software, 21,320 logs).

\subsubsection{Code generation}
Code generation tasks involve automatically synthesising code snippets from natural language descriptions, enabling developers to translate high-level intent into executable code. Recently, 
LLM-based code generator has become mainstream \cite{jiang2024survey, mu2024clarifygpt}, which can reduce manual coding effort and increase productivity. 
In our expriments, we use a curated subset of the CoNaLa dataset~\cite{yin2018learning} consisting of 2,879 examples, focusing on generating Python code from natural language descriptions.

\subsection{Baselines}
\subsubsection{Individual LLMs}
Assigning tasks to a specific LLM is a common approach in practice; therefore, we select individual LLMs as baselines for comparison against the SLS. Specifically, we submit the entire set of queries to each LLM and evaluate their performance in terms of cost, processing time, and success rate. In this paper, 8 candidate LLMs are selected from 3 mainstream providers: OpenAI\textsuperscript{\ref{openai}} (GPT4o, GPT4o-mini, GPT-3, and GPT-4), Anthropic\textsuperscript{\ref{anthropic}} (Claude-3.5-sonnet, Claude-3-opus, Claude-3-haiku), and Meta\textsuperscript{\ref{meta}} (Llama-8B-Turbo , Llama-70B-Turbo).

\subsubsection{Related frameworks}
There are a few existing frameworks for optimizing LLM invocation. To evaluate the effectiveness of SLS, we compare it with OptLLM~\cite{liu2024optllm}, a multi-objective optimization framework using static scheduling. It offers a set of optimal solutions based on budget constraints and performance preferences. Since OptLLM is designed for multi-objective optimization, we record the solution that achieves the highest performance with its corresponding cost and time for comparison.

\subsubsection{Baseline Scheduling Strategies}
To illustrate the effectiveness of SLS’s intelligent scheduling, we also compare it with the following baseline scheduling strategies:
a) First-In-First-Out (FIFO) scheduling~\cite{pan2005fifo} is a classical and straightforward approach to job allocation, where jobs are processed strictly in the order they arrive. 
b) Random selection is also used as a baseline to compare against the SLS, providing a benchmark for non-optimized LLM assignment. We evaluate this method by submitting all queries to one of the candidate LLMs, without considering specific task characteristics or cost-performance metrics.

\subsection{Evaluation metrics}
\label{sec5Evaluation metric}
\noindent\textbf{Total Cost.} This metric represents the overall cost incurred, including the invocation cost of the LLM APIs and the generation cost at the unit of USD.
It evaluates the cost-effectiveness of the solution. 
\begin{equation}
\quad f_{\text{cost}} = \sum_{t=1}^{T} f_{\text{cost}}^t
\end{equation}
\rv{For a fair comparison, the initialization cost is included for all methods that utilize the 1\% training data (i.e., SLS and OptLLM).}

\noindent\textbf{Performance.} This metric measures by the accuracy, namely the percentage of queries that have been processed correctly, indicating the effectiveness of the LLM in delivering accurate results. 
\begin{equation}
\quad f_{\text{perf}} = \frac{1}{\sum_{t=1}^{T} n_t} \sum_{t=1}^{T} \sum_{i \in J_t} perf_i^t
\end{equation}
For log parsing, accuracy is measured as the number of outputs exactly matching the ground truth labels. For code generation, we use BLEU score~\cite{papineni2002bleu} to evaluate the correctness and similarity of generated code to reference solutions.

\noindent\textbf{Timespan.} To further demonstrate the effectiveness of SLS, we report makespan (total completion time) as an additional metric. 
Makespan represents the total time required to complete all queries, including LLM processing time and cache retrieval latency. 
Assuming that queries arrived in one scheduling horizon cannot affect later ones, the total makespan is the sum of the makespan for each interval:
\begin{equation}
f_{\text{makespan}}^t = \max_{l \in L} ( f_{\text{time}, l}^t + f_{\text{cache}}^t )
\end{equation}
Then the makespan can be calculated by:
\begin{equation}
f_{\text{makespan}} = \sum_{t=1}^{T} f_{\text{makespan}}^t 
\end{equation}
\rv{Similarly, the initialization time (i.e., the time required to collect initial training data by submitting jobs to all candidate LLMs for training the prediction model) is included for all methods that utilize the 1\% training data (i.e., SLS and OptLLM).}

\subsection{Implementation}
To mitigate the impact of randomness and ensure robustness in our findings, each experiment is conducted 30 times, and the average values are reported. The initial training set used for initializing the framework consists of 1\% of the dataset. 

The remaining data are used to test the optimization of LLM invocation, with query arrivals modeled by a Poisson distribution. 
The Poisson process is frequently applied to scheduling problems due to its effectiveness in representing random and independent events occurring within fixed time intervals~\cite{qu2024cloud, guo2024outpatient, liang2023poisson}. In this study, the Poisson arrival rate $\lambda$ is determined based on the dataset size. Specifically, for a dataset of size $s$, $\lambda$ is randomly selected from the interval $[0.1s,0.2s]$. For instance, with a dataset size 1000, $\lambda$ would range between 100 and 200. This scaling ensures that the arrival rate adapts proportionally to the dataset size, providing a realistic simulation of varying workloads.

\section{Experimental Results and Analysis}
\label{sec6:Experimental Results and Analysis}

\subsection{RQ1: Comparison with the baselines}
We evaluated the performance of SLS against various baseline solutions across multiple datasets to assess its effectiveness in optimizing LLM invocation. Each experiment is repeated 30 times, and the average results are reported. Tables~\ref{tb_res_log} and \ref{tb_res_code} present the comparison of performance, cost, and makespan for log parsing and code generation tasks, respectively.
\begin{center}

\begin{table}[h]
\caption{Overall performance on the log parsing task}
\vspace{-6pt}
\vspace{-6pt}
\label{tb_res_log}
\resizebox{1\columnwidth}{!}{%
\begin{tabular}{clllllll}
\midrule[1pt]
Dataset & Solution & \multicolumn{2}{c}{Performance} & \multicolumn{2}{c}{Cost} & \multicolumn{2}{c}{Makespan} \\
\multicolumn{1}{l}{} &  & Avg & IMPV  & Avg & SAVING  & Avg & SAVING  \\ \midrule[1pt]
Apache & GPT4o & 0.9910 & 0.59\% & 33.95 & 92.87\% & 868 & 96.04\% \\
 & GPT4o-mini & 0.9427 & 5.74\% & \textbf{1.19} & -103.16\% & 963 & 96.43\% \\
 & GPT-3.5-turbo & 0.8355 & 19.31\% & 3.53 & 31.38\% & 908 & 96.21\% \\
 & Claude-3-5-sonnet & 0.9702 & 2.74\% & 38.99 & 93.79\% & 963 & 96.43\% \\
 & Claude-3-opus & 0.9954 & 0.14\% & 150.17 & 98.39\% & 1926 & 98.21\% \\
 & Claude-3-haiku & 0.8709 & 14.46\% & 2.58 & 5.99\% & 616 & 94.41\% \\
 & Llama-8B-Turbo & 0.7948 & 25.42\% & 1.21 & -100.92\% & 1078 & 96.81\% \\
 & Llama-70B-Turbo & 0.8800 & 13.27\% & 5.73 & 57.74\% & 1267 & 97.29\% \\
 & FIFO & 0.9050 & 10.14\% & 21.06 & 88.50\% & 122 & 71.80\% \\
 & Random Selection & 0.9098 & 9.56\% & 29.67 & 91.84\% & 1074 & 96.80\% \\
 & OptLLM & 0.9961 & 0.07\% & 74.52 & 96.75\% & 860 & 96.00\% \\
 & SLS & \textbf{0.9968} & \textbackslash{} & 2.42 & \textbackslash{} & \textbf{34} & \textbackslash{} \\  \hline
Linux & GPT4o & 0.6078 & 51.09\% & 16.50 & 90.86\% & 481 & 91.53\% \\
 & GPT4o-mini & 0.5655 & 62.39\% & 0.60 & -150.65\% & 400 & 89.82\% \\
 & GPT-3.5-turbo & 0.8041 & 14.20\% & 1.70 & 11.04\% & 454 & 91.02\% \\
 & Claude-3-5-sonnet & 0.6567 & 39.84\% & 17.77 & 91.52\% & 444 & 90.83\% \\
 & Claude-3-opus & 0.1840 & 399.08\% & 83.20 & 98.19\% & 944 & 95.68\% \\
 & Claude-3-haiku & 0.6481 & 41.69\% & 1.15 & -31.27\% & 294 & 86.16\% \\
 & Llama-8B-Turbo & 0.5525 & 66.21\% & \textbf{0.58} & -160.43\% & 359 & 88.65\% \\
 & Llama-70B-Turbo & 0.6426 & 42.90\% & 2.67 & 43.62\% & 1285 & 96.83\% \\
 & FIFO & 0.6075 & 51.16\% & 10.28 & 85.33\% & 59 & 30.93\% \\
 & Random Selection & 0.5831 & 57.49\% & 15.50 & 90.27\% & 582 & 93.00\% \\
 & OptLLM & 0.8599 & 6.80\% & 8.28 & 81.79\% & 220 & 81.48\% \\
 & SLS & \textbf{0.9183} & \textbackslash{} & 1.51 &\textbackslash{}  & \textbf{41} & \textbackslash{} \\  \hline
Proxifier & GPT4o & 0.4550 & 65.82\% & 19.01 & 50.08\% & 470 & 80.56\% \\
 & GPT4o-mini & 0.1788 & 321.98\% & 0.64 & -1373.36\% & 493 & 81.46\% \\
 & GPT-3.5-turbo & 0.4149 & 81.85\% & 1.88 & -404.19\% & 434 & 78.96\% \\
 & Claude-3-5-sonnet & 0.3896 & 93.66\% & 20.84 & 54.45\% & 386 & 76.34\% \\
 & Claude-3-opus & 0.7081 & 6.55\% & 101.49 & 90.65\% & 805 & 88.64\% \\
 & Claude-3-haiku & 0.0252 & 2894.05\% & 1.49 & -536.99\% & 261 & 64.95\% \\
 & Llama-8B-Turbo & 0.0307 & 2357.65\% & \textbf{0.59} & -1504.38\% & 411 & 77.77\% \\
 & Llama-70B-Turbo & 0.3622 & 108.31\% & 2.74 & -245.92\% & 859 & 89.36\% \\
 & FIFO & 0.2689 & 180.59\% & 13.34 & 28.86\% & \textbf{56} & -63.24\% \\
 & Random Selection & 0.3200 & 135.78\% & 18.58 & 48.92\% & 515 & 82.25\% \\
 & OptLLM & 0.7072 & 6.69\% & 96.30 & 90.15\% & 760 & 87.97\% \\
 & SLS & \textbf{0.7545} & \textbackslash{}& 9.49 & \textbackslash{}  & 91 & \textbackslash{} \\ 
 \midrule[1pt]
\end{tabular}%
}
\vspace{-6pt}
\end{table}

\end{center}

\begin{center}


\begin{table}[ht]
\caption{Overall performance on the code generation}
\vspace{-6pt}
\label{tb_res_code}
\resizebox{1\columnwidth}{!}{%
\begin{tabular}{llllllll}
\midrule[1pt]
\multicolumn{1}{c}{Dataset} & Solution & \multicolumn{2}{c}{Performance} & \multicolumn{2}{c}{Cost} & \multicolumn{2}{c}{Makespan} \\
 &  & Avg & IMPV  & Avg & SAVING  & Avg & SAVING  \\ \midrule[1pt]
\multirow{10}{*}{CoNaLa} & GPT-4o & 0.1085 & 47.65\% & 1.77 & 54.55\% & 50 & 57.92\% \\
 & GPT-4o-mini & 0.0810 & 95.93\% & \textbf{0.07} & -985.33\%& 50 & 57.98\% \\
 & GPT-3.5-turbo & 0.0987 & 60.79\% & 0.18 & -353.80\% & 51 & 58.88\% \\
 & Claude-3-5-sonnet & 0.1487 & 6.72\% & 1.46 & 44.65\% & 56 & 62.39\% \\
 & Claude-3-opus & 0.1226 & 29.45\% & 9.27 & 91.31\% & 121 & 82.64\% \\
 & Claude-3-haiku & 0.1126 & 40.94\% & 0.16 & -413.30\% & 38 & 44.72\% \\
 & Llama-8B-Turbo & 0.0060 & 2545.00\% & 0.09 & -781.31\% & 76 & 72.33\% \\
 & Llama-70B-Turbo & 0.0054 & 2838.89\% & 0.49 & -63.80\% & 258 & 91.86\% \\
 & FIFO & 0.0967 & 64.12\% & 1.19 & 32.24\% & \textbf{8} & -162.71\% \\
 & Random Selection & 0.0853 & 86.05\% & 1.69 & 52.28\% & 32 & 34.32\% \\
 & {OptLLM}  & 0.1217 & 30.40\% & 2.44 & 66.94\% & 31 & 32.11\%\\
& {SLS} & {\textbf{0.1587}} & \textbackslash{} & {0.81} & \textbackslash{} & {21} &\textbackslash{}  \\ 
 \midrule[1pt]
\end{tabular}%
}
\end{table}
\end{center}

\begin{figure}
    \centering
\includegraphics[width=1\linewidth]{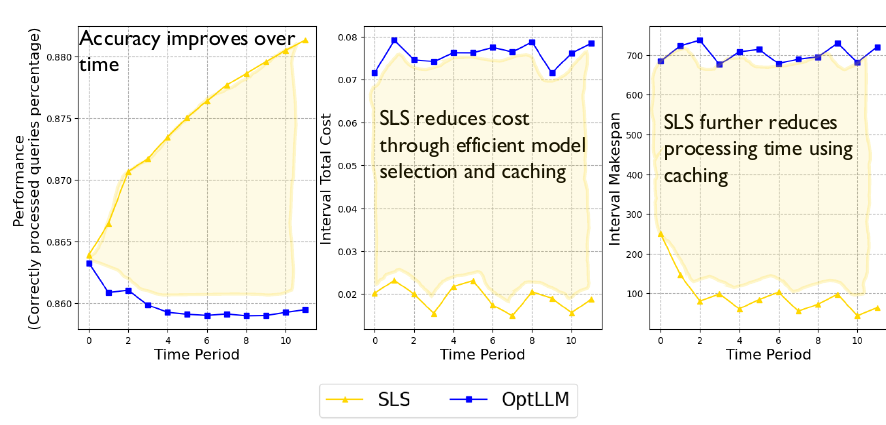}
    \caption{Interval Metrics Comparison between SLS and OptLLM on a Single Run of the Linux Dataset (Log Parsing)}
    \label{fig:slm_benefit}
    \vspace{-12pt}
\end{figure}

\vspace{-30pt}
Table~\ref{tb_res_log} summarizes comparison results for log parsing across three datasets: Apache, Linux, and Proxifier. The results indicate that SLS consistently outperforms the baselines in performance while significantly reducing cost and makespan. For example, on the Apache dataset, SLS achieved the highest average performance \rv{(0.9968)}, exceeding both the best individual LLM, Claude-3-opus (0.9954), and the OptLLM framework (0.9961). Moreover, SLS delivered this performance at a substantially lower cost of \$2.42, compared to \$150.17 for Claude-3-opus and \$74.52 for OptLLM. Furthermore, SLS reduced makespan from 71.80\% compared to baseline methods. The significant cost and makespan savings can be attributed to the effective utilization of cached results, as identical queries frequently appear in the log parsing task. Similar trends are observed in the Linux and Proxifier datasets, where SLS consistently delivered superior performance and efficiency.

The results for code generation are summarized in Table~\ref{tb_res_code}. The table shows that the candidate LLMs exhibit varying performance levels for the code generation task. Claude-3-5-sonnet achieves the highest performance among individual models, with an accuracy of 0.1487 at \$1.46. In contrast, Lllama-8B-Turbo performs poorly, with an accuracy of only 0.0060, despite its low cost of \$0.09. SLS outperforms all individual models, achieving an accuracy of \rv{0.1587} at a significantly lower cost of \rv{\$0.81}. It also achieves the second-shortest makespan, completing all jobs in 21 minutes. Compared to the log parsing task, SLS achieves limited cost savings in this scenario. Two potential reasons may account for this. First, all candidate LLMs perform poorly on this task, with each individual model processing only around 10\% of jobs. The cheapest LLM, GPT-4o-mini, achieves an accuracy of only 0.0810. Consequently, SLS prioritizes LLMs with a higher success rate, even though they come at a higher cost. Second, the jobs in CoNaLa exhibit significant variation, which reduces the effectiveness of the Cache. As a result, SLS benefits less from cache hits, further limiting cost savings. Nevertheless, SLS still improves overall performance and reduces processing time by efficiently balancing the trade-off between accuracy and cost. 

Figure~\ref{fig:slm_benefit} illustrates the performance benefits of SLS’s dynamic scheduling compared to the static allocation strategy used in OptLLM. \rv{It presents the period-by-period performance change from a randomly selected run on the Linux dataset (i.e., one of the 30 experimental runs).} By continuously monitoring and updating both the cache and the prediction model, SLS progressively improves query processing accuracy. It reduces interval costs through effective model selection and caching, and further optimizes processing time via its efficient caching strategy.

\rv{In summary, SLS achieves the highest performance across all datasets, with improvements ranging from 0.07\% to 2894.05\%. While it does not always achieve the lowest cost or makespan, it consistently reduces both metrics in most cases. Specifically, SLS achieves cost savings between 3.68\% and 98.39\%, and processing time reductions between 22.58\% and 98.21\% compared to higher-cost or slower baselines.}
To validate the comparison, we perform a Friedman test, a non-parametric statistical method suitable for comparing multiple models across repeated measurements. We set the significance level ($\alpha$) to 0.05, and the results confirm statistically significant differences among the methods. 
\rv{To further identify where these differences lie, we apply a post-hoc Nemenyi test, which compares all pairs of models to determine whether their performance differences are statistically significant. }
Overall, the experimental results show that SLS consistently achieves the highest performance across all evaluated datasets while reducing cost and makespan.

\subsection{RQ2: Contribution and impact of SLS components}
We conducted an ablation study to evaluate the contribution of each optimization component in SLS by designing three variant algorithms. SLS$_{\text{w/o c}}$ excludes the Cache Manager, preventing the reuse of stored results. SLS$_{\text{w/o s}}$ removes the Scheduler, allocating jobs randomly instead of using optimized scheduling. SLS$_{\text{w/o u}}$ disables the Update Manager, keeping the prediction model and cache configurations unchanged throughout. The comparison of these variants against the complete SLS is presented in Table~\ref{tb_add_ablation}.

\begin{center}
\begin{table}[htbp]
\caption{Ablation study results
}
\vspace{-6pt}
\label{tb_add_ablation}
\resizebox{1\columnwidth}{!}{%
\begin{tabular}{lllllllllll}
\midrule[1pt]
\multirow{2}{*}{Dataset} & \multicolumn{1}{c}{\multirow{2}{*}{Solution}} & \multicolumn{3}{c}{Performance} & \multicolumn{3}{c}{Total Cost} & \multicolumn{3}{c}{Makespan} \\
 & \multicolumn{1}{c}{} & \multicolumn{1}{c}{Avg} & \multicolumn{1}{c}{Max} & \multicolumn{1}{c}{Min} & \multicolumn{1}{c}{Avg} & \multicolumn{1}{c}{Max} & \multicolumn{1}{c}{Min} & \multicolumn{1}{c}{Avg} & \multicolumn{1}{c}{Max} & \multicolumn{1}{c}{Min} \\  \midrule[1pt]
\multirow{4}{*}{Apache} & SLS  & \textbf{0.9968} & \textbf{0.9981} & \textbf{0.9949} & \textbf{2.42} & \textbf{2.48} & \textbf{2.40} & 34 & 85 & 26 \\
 & $\text{SLS}_{\text{w/o c}}$ & 0.9958 & 0.9974 & 0.9942 & 3.68 & 3.83 & 3.61 & 639 & 792 & 485 \\
 & $\text{SLS}_{\text{w/o s}}$ & 0.9952 & 0.9965 & 0.9941 & 3.19 & 3.36 & 2.93 & \textbf{26} & \textbf{28} & \textbf{24} \\
 & $\text{SLS}_{\text{w/o u}}$ & 0.9936 & 0.9965 & 0.9893 & 2.49 & 2.52 & 2.47 & 92 & 105 & 80 \\ \hline
\multirow{4}{*}{Linux} & SLS & \textbf{0.9183} & \textbf{0.9303} & \textbf{0.9051} & 1.51 & 1.59 & 1.44 & 41 & 47 & 33 \\
 & $\text{SLS}_{\text{w/o c}}$ & 0.8595 & 0.8634 & 0.8536 & 2.66 & 2.82 & 2.51 & 283 & 348 & 200 \\
 & $\text{SLS}_{\text{w/o s}}$ & 0.9072 & 0.9172 & 0.8878 & 2.45 & 2.83 & 2.16 & \textbf{23} & \textbf{26} & \textbf{20} \\
 & $\text{SLS}_{\text{w/o u}}$ & 0.9133 & 0.9274 & 0.8933 & \textbf{1.45} & \textbf{1.51} & \textbf{1.42} & 52 & 61 & 38 \\ \hline
\multirow{4}{*}{Proxifier} & SLS & \textbf{0.7545} & \textbf{0.7916} & \textbf{0.6959} & 9.49 & 13.70 & 7.28 & 91 & 138 & 71 \\
 & $\text{SLS}_{\text{w/o c}}$ & 0.5672 & 0.5896 & 0.5425 & 10.39 & 13.13 & 7.93 & 232 & 273 & 176 \\
 & $\text{SLS}_{\text{w/o s}}$ & 0.6014 & 0.6971 & 0.4874 & 8.08 & 12.52 & 4.70 & \textbf{46} & \textbf{70} & \textbf{27} \\
 & $\text{SLS}_{\text{w/o u}}$ & 0.6031 & 0.6312 & 0.5706 & \textbf{5.50} & \textbf{9.91} & \textbf{2.75} & 162 & 270 & 101 \\ \hline
\multirow{4}{*}{CoNaLa} & SLS & \textbf{0.1587} & \textbf{0.1597} & \textbf{0.1574} & \textbf{0.81} & \textbf{0.87} & \textbf{0.78} & \textbf{21} & \textbf{22} & \textbf{20} \\
 & $\text{SLS}_{\text{w/o c}}$ & \textbf{0.1587} & \textbf{0.1597} & \textbf{0.1574} & \textbf{0.81} & \textbf{0.87} & \textbf{0.78} & \textbf{21} & \textbf{22} & \textbf{20} \\
 & $\text{SLS}_{\text{w/o s}}$ & 0.0853 & 0.0901 & 0.0820 & 1.82 & 1.94 & 1.67 & 35 & 40 & 31 \\
 & $\text{SLS}_{\text{w/o u}}$ & 0.1557 & 0.1577 & 0.1537 & 0.81 & 0.87 & 0.72 & 20 & 22 & 18 \\  
 \hline
 \multirow{4}{*}{Headlines} & SLS & \textbf{0.7890} & \textbf{0.8000} & \textbf{0.7709} & \textbf{0.50} & \textbf{0.64} & \textbf{0.36} & 50 & 84 & \textbf{33} \\
 & $\text{SLS}_{\text{w/o c}}$ & 0.7602 & 0.7683 & 0.7455 & 0.72 & 0.89 & 0.49 & 86 & 130 & 45 \\
 & $\text{SLS}_{\text{w/o s}}$ & 0.7432 & 0.7464 & 0.7401 & 2.88 & 2.99 & 2.79 & \textbf{39} & \textbf{47} & 35 \\
 & $\text{SLS}_{\text{w/o u}}$ & 0.7574 & 0.7720 & 0.7344 & 0.71 & 0.91 & 0.42 & 91 & 145 & 47 \\ \hline
\multirow{4}{*}{Overruling} & SLS & \textbf{0.9564} & \textbf{0.9599} & \textbf{0.9545} & 1.10 & 1.27 & 1.00 & 24 & \textbf{29} & 23 \\
 & $\text{SLS}_{\text{w/o c}}$ & 0.9453 & 0.9511 & 0.9389 & \textbf{0.46} & \textbf{0.56} & 0.31 & 29 & 53 & \textbf{18} \\
 & $\text{SLS}_{\text{w/o s}}$ & 0.9364 & 0.9414 & 0.9313 & 2.96 & 3.19 & 2.68 & \textbf{22} & \textbf{29} & 19 \\
 & $\text{SLS}_{\text{w/o u}}$ & 0.9448 & 0.9515 & 0.9359 & \textbf{0.46} & 0.57 & \textbf{0.29} & 31 & 53 & \textbf{18} \\
 \midrule[1pt]
\end{tabular}
}
\end{table}

\end{center}
\vspace{-12pt}
\textbf{Impact of the Cache Manager.}
The experimental results show that removing the Cache Manager leads to a significant increase in both total cost and processing time, particularly on datasets with identical or highly similar jobs, while the impact on performance varies across datasets. 
For instance, on the Apache dataset, the total cost increases from \$2.42 to \$3.68, and processing time rises from 34 to 643 minutes, while performance only slightly decreases from 0.9968 to 0.9958. Although Apache contains a considerable number of identical jobs, the candidate LLMs demonstrate relatively stable performance across jobs, allowing the Scheduler to maintain high accuracy even without cache support. However, the absence of caching leads to increased cost and makespan, as all jobs must be submitted to LLMs without reuse. 
However, on the Linux and Proxifier datasets, performance is more sensitive. Without the Cache Manager, all jobs are routed through the Scheduler. If the prediction model fails to select the correct LLM, performance drops significantly.
In contrast, the CoNaLa dataset shows minimal differences with or without cache. This is because CoNaLa contains no identical or highly similar jobs, inherently limiting the effectiveness of cache reuse. As a result, the performance and efficiency of SLS remain largely unchanged in the absence of the Cache Manager.

\textbf{Impact of the Scheduler.}
Excluding the Scheduler led to significant performance degradation, particularly in tasks where the performance of LLMs varies considerably. \rv{For example, the average performance on the Proxifier dataset decreased from 0.7545 to 0.6014\%, and the total cost increased significantly from \$0.27 to \$2.76.} This highlights the Scheduler’s critical role in robust performance by selecting the most suitable LLM for each query, reducing the likelihood of suboptimal model choices.

\textbf{Impact of the Update Manager.}
Disabling the Update Manager caused a slight performance degradation and increased cost and time variability. On the Linux dataset, for instance, performance dropped from 0.9183 to 0.9133 and makespan increased. Although the impact is less significant than removing the Cache Manager or Scheduler, it suggests that the Update Manager is essential in maintaining model accuracy and efficiency over time.

\subsection{RQ3: The generalizability of SLS}
To ensure that the proposed schedule optimization method is not limited to a particular task, we extend our study to an NLP task - text classification. Text classification involves assigning predefined labels to given text inputs, such as sentiment polarity or topic categories. For example, sentiment analysis tasks analyze the sentiment expressed in text data and identify whether it is positive, negative, or neutral. We use the Headlines dataset~\cite{sinha2021impact}, which contains 10,000 news headlines for gold price changes, and the Overruling dataset~\cite{zheng2021does}, which consists of 2,400 legal documents labeled according to whether a court decision was overruled.

Table~\ref{tb_res_text} presents the results for text classification on the Headlines and Overruling datasets. SLS achieved the highest average performance on both datasets, with an accuracy of \rv{0.7890 for Headlines and 0.9564 for Overruling. These results surpass the baselines, with performance improvements ranging from 0.38\% to 11.71\% on Headlines and 0.63\% to 6.6\% on Overruling. Similarly, SLS achieved significant cost savings, reducing costs by up to 96.1\% on Headlines and 93.04\% on Overruling. In terms of makespan, it reduced the total completion time by 53.70\% to 82.58\% on Headlines except FIFO, and 22.58\% to 84.42\% on Overruling.} In addition to cache reuse, the Scheduler enables the parallel invocation of multiple LLMs, further reducing makespan. The results demonstrate that the SLS generalizes well, performing effectively not only on domain-specific software engineering tasks but also on general NLP tasks.

\begin{center}
\begin{table}[ht]
\caption{Overall performance on the text classification}
\vspace{-6pt}
\label{tb_res_text}
\resizebox{1\columnwidth}{!}{%
\begin{tabular}{lllllllllll}
\midrule[1pt]
\multicolumn{1}{c}{Dataset} & Solution & \multicolumn{2}{c}{Performance} & \multicolumn{2}{c}{Cost} & \multicolumn{2}{c}{Makespan} \\
 &  & AVG & IMPV & AVG & SAVING & AVG & SAVING \\ \midrule[1pt]
\multirow{12}{*}{Headlines} & GPT-4o & 0.7515 & 4.99\% & 4.01 & 87.43\% & 152 & 67.11\% \\
 & GPT-4o-mini & 0.7317 & 7.83\% & \textbf{0.12} & -314.05\% & 135 & 62.96\% \\
 & GPT-3.5-turbo & 0.7684 & 2.68\% & 0.41 & -23.99\% & 147 & 65.99\% \\
 & Claude-3-5-sonnet & 0.7419 & 6.35\% & 2.73 & 81.57\% & 150 & 66.67\% \\
 & Claude-3-opus & 0.7861 & 0.38\% & 12.93 & 96.10\% & 287 & 82.58\% \\
 & Claude-3-haiku & 0.7254 & 8.77\% & 0.23 & -123.16\% & 136 & 63.24\% \\
 & Llama-8B-Turbo & 0.7063 & 11.71\% & 0.19 & -169.03\% & 126 & 60.32\% \\
 & Llama-70B-Turbo & 0.7463 & 5.72\% & 0.92 & 45.49\% & 153 & 67.32\% \\
 & FIFO & 0.7412 & 9.40\% & 1.95 & 74.16\% & \textbf{20} & -150.00\% \\
 & Random Selection & 0.7448 & 5.93\% & 2.69 & 81.27\% & 160 & 68.75\% \\
 & OptLLM & 0.7734 & 2.02\% & 5.23 & 90.37\% & 108 & 53.70\% \\
 & SLS & \textbf{0.7890} & \textbackslash{} & 0.50 & \textbackslash{} & 50 & \textbackslash{} \\ \hline
\multirow{10}{*}{Overruling} & GPT-4o & 0.8971 & 6.60\% & 1.36 & 19.26\% & 35 & 31.43\% \\
 & GPT-4o-mini & 0.9461 & 1.08\% & \textbf{0.04} & -2645.25\% & 31 & 22.58\% \\
 & GPT-3.5-turbo & 0.9379 & 1.98\% & 0.14 & -684.36\% & 65 & 62.50\% \\
 & Claude-3-5-sonnet & 0.9499 & 0.68\% & 3.67 & 70.08\% & 81 & 70.37\% \\
 & Claude-3-opus & 0.9494 & 0.73\% & 15.77 & 93.04\% & 154 & 84.42\% \\
 & Claude-3-haiku & 0.9434 & 1.42\% & 0.22 & -399.14\% & 45 & 46.67\% \\
 & Llama-8B-Turbo & 0.9177 & 4.18\% & 0.06 & -1730.17\% & 31 & 22.58\% \\
 & Llama-70B-Turbo & 0.9505 & 0.63\% & 0.3 & -266.03\% & 35 & 31.43\% \\
 & FIFO & 0.9337 & 2.43\% & 1.14 & 3.68\% & \textbf{7} & 31.43\% \\
 & Random Selection & 0.9363 & 2.15\% & 2.7 & 59.33\% & 60 & 60.00\% \\
 & OptLLM & 0.9445 & 1.26\% & 5.19 & 78.84\% & 53 & 54.72\% \\
 & SLS & \textbf{0.9564} & \textbackslash{} & 1.10 & \textbackslash{} & 24 & \textbackslash{} \\ 
 \midrule[1pt]
\end{tabular}%
}
\vspace{-6pt}
\end{table}

\end{center}

\subsection{RQ4: Effect of different parameter settings}

\subsubsection{Cache similarity threshold $\tau$}
\rv{
The similarity threshold $\tau$ in the cache controls how strictly the system matches new queries to cached entries based on embedding similarity. To analyze the impact of this parameter, we tested SLS using four different values of $\tau$: 0.8, 0.9, and 0.99. To focus on the effect of $\tau$, we disabled the cache update mechanism during this evaluation, ensuring that the threshold remained fixed throughout the process. In addition to reporting the average performance, cost, and makespan over 30 runs, we also present the average values of the periodical metrics, including cache hit rate, cache performance, scheduler hit rate, and scheduler performance. Specifically, for each run, we compute the average of the metric across all periods, and then report the overall average across the 30 runs. The results are given in Table~\ref{tb_res_para_tau}.

As shown in the table, $\tau$ has a significant impact on overall system performance, with the exception of CoNaLa. If $\tau$ is set too low, the cache may return many false matches, increasing the cache hit rate but lowering cache correctness. This can negatively affect SLS’s overall performance. To maintain high cache correctness, a higher threshold is required; however, this may reduce the number of cache hits and thereby reduce the time and cost benefits of caching. In contrast, for the CoNaLa dataset, the results remain unchanged regardless of the $\tau$ value. A closer examination reveals that the cache hit rate remains consistently zero for CoNaLa, indicating the absence of identical or semantically similar queries. As a result, the cache provides minimal benefit, and variations in the cache parameter have little to no effect on performance for this dataset.
}
\subsubsection{Update frequency control parameter $Q$}
The Update frequency control parameter $Q$ controls the frequency of adaptive updates within the SLS framework, as defined in Equation~\ref{mod}. To identify the impact of update frequency, we tested $Q$ with values of 1, 3, and 5. The average results are given in Table~\ref{tb_new_res_para}. A smaller $Q$ slightly improves performance, as more frequent checks allow SLS to evaluate the need for updates more regularly. However, the performance gains are minimal, as the conditions for triggering updates are not always met, even with frequent evaluations. Setting $Q$ too low could also lead to overfitting due to excessive updates.

\begin{center}

\begin{table}[]
\caption{The performance of SLS with different settings on cache similarity threshold $\tau$}
\vspace{-6pt}
\label{tb_res_para_tau}
\resizebox{1\columnwidth}{!}{%
\begin{tabular}{ccccccccc}
\midrule[1pt]
\multirow{2}{*}{Dataset} & \multirow{2}{*}{$\tau$} & \multicolumn{3}{c}{Total} & \multicolumn{2}{c}{Cache} & \multicolumn{2}{c}{Scheduler} \\
 &  & Perf. & Cost & Makespan & \multicolumn{1}{c}{Hit Rate} & \multicolumn{1}{c}{Perf.} & \multicolumn{1}{c}{Hit Rate} & \multicolumn{1}{c}{Perf.} \\ \midrule[1pt]
\multirow{3}{*}{Apache} & 0.8 & 0.9971 & \textbf{2.41} & \textbf{28} & 98.57\% & 0.9988 & 1.43\% & 0.8823 \\
 & 0.9 & \textbf{0.9980} & 2.43 & 41 & 97.21\% & 0.9998 & 2.79\% & 0.9404 \\
 & 0.99 & 0.9977 & 3.07 & 413 & 46.07\% & 1.0000 & 53.93\% & 0.9957 \\ \hline
\multirow{3}{*}{Linux} & 0.8 & 0.9209 & \textbf{1.55} & \textbf{68} & 81.90\% & 0.9895 & 18.10\% & 0.6095 \\
 & 0.9 & \textbf{0.9225} & 1.68 & 95 & 74.48\% & 0.9988 & 25.52\% & 0.6997 \\
 & 0.99 & 0.9011 & 2.17 & 198 & 30.77\% & 1.0000 & 69.23\% & 0.8569 \\ \hline
\multirow{3}{*}{Proxifier} & 0.8 & \textbf{0.7294} & \textbf{3.77} & \textbf{98} & 68.44\% & 0.7985 & 31.56\% & 0.4658 \\
 & 0.9 & 0.7027 & 4.80 & 116 & 54.06\% & 0.7548 & 45.94\% & 0.4970 \\
 & 0.99 & 0.6655 & 6.42 & 170 & 36.95\% & 0.9333 & 63.05\% & 0.4672 \\ \hline
\multirow{3}{*}{CoNaLa} & 0.8 & \textbf{0.1587} & \textbf{0.81} & \textbf{21} & 0.00\% & 0.0000 & 100\% & 0.1587 \\
 & 0.9 & \textbf{0.1587} & \textbf{0.81} & \textbf{21} & 0.00\% & 0.0000 & 100\% & 0.1587 \\
 & 0.99 & \textbf{0.1587} & \textbf{0.81} & \textbf{21} & 0.00\% & 0.0000 & 100\% & 0.1587 \\ \hline
\multirow{3}{*}{Headlines} & 0.8 & 0.7785 & \textbf{0.69} & \textbf{83} & 5.87\% & \multicolumn{1}{l}{0.6742} & 94.13\% & \multicolumn{1}{l}{0.7699} \\
 & 0.9 & \textbf{0.7887} & 0.71 & 89 & 0.63\% & \multicolumn{1}{l}{0.7109} & 99.37\% & \multicolumn{1}{l}{0.7783} \\
 & 0.99 & 0.7690 & 0.70 & 87 & 0.00\% & 0.0000 & 100.00\% & 0.7690 \\ \hline
\multirow{3}{*}{Overruling} & 0.8 & 0.9558 & \textbf{1.09} & \textbf{24} & 0.02\% & 0.0292 & 99.98\% & 0.9557 \\
 & 0.9 & \textbf{0.9564} & 1.10 & \textbf{24} & 0.00\% & 0.0000 & 100.00\% & 0.9564 \\
 & 0.99 & 0.9563 & 1.10 & \textbf{24} & 0.00\% & 0.0000 & 100.00\% & 0.9563 \\
 \midrule[1pt]
\end{tabular}%
}
\vspace{-12pt}
\end{table}
\end{center}

\begin{center}
\begin{table}[]
\small
\caption{The performance of SLS with different parameter settings}
\vspace{-6pt}
\label{tb_new_res_para}
\resizebox{0.95\columnwidth}{!}{%
\begin{tabular}{l|cccc|cccc}
\hline
Dataset & Q & Perf & Cost & Makespan & Ratio & Perf & Cost & Makespan\\ \hline
Apache & 1 & \textbf{0.9968} & \textbf{2.42} & \textbf{34} & 2 & \textbf{0.9977} & 3.07 & 413 \\
 & 3 & 0.9950 & 2.49 & 76 & 1 & 0.9968 & 2.42 & 34 \\
 & 5 & 0.9955 & 2.43 & 38 & 0.5 & 0.9971 & \textbf{2.41} & \textbf{28} \\ \hline
\multirow{3}{*}{Linux} & 1 & \textbf{0.9183} & 1.51 & 41 & 2 & \textbf{0.9230} & 2.17 & 198 \\
 & 3 & 0.9133 & \textbf{1.50} & \textbf{40} & 1 & 0.9183 & \textbf{1.51} & \textbf{41} \\
 & 5 & 0.9131 & 1.50 & 41 & 0.5 & 0.9011 & 1.55 & 68 \\ \hline
\multirow{3}{*}{Proxifier} & 1 & \textbf{0.7545} & \textbf{9.49} & \textbf{91} & 2 & 0.7284 & 9.97 & 99 \\
 & 3 & 0.6623 & 12.56 & 129 & 1 & \textbf{0.7545} & 9.49 & \textbf{91} \\
 & 5 & 0.6557 & 13.22 & 132 & 0.5 & 0.6248 & \textbf{2.40} & 148 \\ \hline
\multirow{3}{*}{CoNaLa} & 1 & \textbf{0.1587} & \textbf{0.81} & \textbf{21} & 2 & \textbf{0.1796} & 0.89 & 17 \\
 & 3 & \textbf{0.1587} & \textbf{0.81} & \textbf{21} & 1 & 0.1589 & 0.81 & \textbf{21} \\
 & 5 & \textbf{0.1587} & \textbf{0.81} & \textbf{21} & 0.5 & 0.1415 & \textbf{0.78} & 23 \\ \hline
\multirow{3}{*}{Headlines} & 1 & \textbf{0.7890} & 0.50 & \textbf{50} & 2 & \textbf{0.7913} & 0.60 & \textit{50} \\
 & 3 & 0.7881 & 0.50 & 54 & 1 & 0.7890 & 0.50 & \textit{50} \\
 & 5 & 0.7882 & 0.50 & 53 & 0.5 & 0.7887 & \textbf{0.48} & \textbf{50} \\ \hline
\multirow{3}{*}{Overruling} & 1 & 0.9564 & 1.10 & \textbf{24} & 2 & \textbf{0.9566} & \textbf{1.09} & \textbf{24} \\
 & 3 & 0.9562 & \textbf{1.09} & \textbf{24} & 1 & 0.9564 & 1.10 & \textbf{24} \\
 & 5 & \textbf{0.9565} & \textbf{1.09} & \textbf{24} & 0.5 & 0.9560 & \textbf{1.09} & \textbf{24} \\ \midrule[1pt]
\end{tabular}%
}
\vspace{-6pt}
\end{table}
\end{center}

\subsubsection{Performance-to-Cost ratio parameter}
In the default scheduling rule, we set the performance-to-cost weight to 1, indicating that performance and cost are equally important. This setting aims to maximize the performance per unit cost. To explore the effect of different weightings, we tested the scheduler with adjusted weights.

For instance, setting the weight to \(2\) (i.e., ratio = \( \frac{\text{perf}}{{\text{cost}}^2} \)) prioritizes solutions with higher performance potential, while setting the weight to \(0.5\) (i.e., ratio = \( \frac{\text{perf}}{{\text{cost}}^{0.5}} \)) prefer cheaper solutions. 
As shown in the Table~\ref{tb_new_res_para}, these weight adjustments allow the scheduler to find more suitable solutions depending on the desired trade-off between performance and cost.

\section{Discussion}
\label{sec6discussion}
\subsection{Why Does SLS Work}

There are several factors contributing to the performance of SLS.

\textbf{Efficient Cache Utilization.} By leveraging the cache, SLS significantly reduces makespan and cost by reusing stored results. The time required for cache retrieval is considerably less than that for invoking an LLM, especially for repeated/similar queries.

\textbf{Parallel Processing of LLMs.} The parallel utilization of multiple LLMs allows SLS to distribute queries efficiently, reducing the need to queue on a single LLM. This capability minimizes waiting time and enhances overall system throughput.

\textbf{Adaptive LLM Selection.} The adaptive scheduler optimizes the LLM selection process, choosing more cost-effective LLMs for less complex queries. This reduces costs and mitigates the risk of suboptimal performance, as SLS can select alternative LLMs to handle cases where other models performs poorly. For instance, on the Proxifier dataset, some LLMs exhibited significantly lower performance (e.g., Claude-3-haiku’s accuracy was only 0.0252), while SLS’s adaptive scheduling strategy avoided these underperforming models, achieving an accuracy of 0.7545.

\rv{
\subsection{Lightweight Inspection for Prediction Model Updates}
To enable dynamic optimization, SLS periodically leverages information from recently processed jobs to refine its components. This requires limited access to ground truth to determine whether outputs are correct or not. SLS adopts a lightweight inspection mechanism, where only a small subset of model outputs are selectively checked. These inspected results provide feedback for both cache expansion and prediction model updates.

To evaluate the cost of this process, we measure how frequently inspections are triggered. The experiments are conducted with the update frequency control parameter $Q=1$, meaning that SLS triggers the update process in every period. Metrics are first computed over all periods within each run, and then averaged across 30 runs. The results are given in Table~\ref{tb_res_inspection}.

The inspection rate indicates the proportion of jobs requiring ground-truth verification. On average, around 1\% of processed jobs are inspected in each period when updates are triggered. 
This design allows the framework to operate effectively with minimal reliance on ground-truth labels, making it suitable for deployment in realistic settings where annotations are costly or limited.

\subsection{Flexibility of SLS Framework}
\begin{center}
\begin{table}[]
\caption{Summary of periodical job numbers and inspection times across datasets}
\vspace{-6pt}
\label{tb_res_inspection}
\resizebox{1\columnwidth}{!}{%
\begin{tabular}{cccccccccccccc}
\midrule[1pt]
\multirow{2}{*}{Dataset} & \multirow{2}{*}{Periods} & \multicolumn{4}{c}{Periodical Job Num} & \multicolumn{4}{c}{Periodical Inspection Times} & \multicolumn{4}{c}{Inspection Rate} \\ \cline{3-14} 
 &  & Avg & Max & Min & Sum & Avg & Max & Min & Sum & Avg & Max & Min & Sum \\ \midrule[1pt]
Apache & 12 & 4102 & 5628 & 3863 & 51459 & 12 & 21 & 4 & 152 & 0.30\% & 0.50\% & 0.10\% & 0.29\% \\
Linux & 13 & 1772 & 2502 & 1643 & 23682 & 10 & 19 & 6 & 139 & 0.59\% & 1.09\% & 0.31\% & 0.59\% \\
Proxifier & 12 & 1648 & 2453 & 1512 & 21107 & 17 & 26 & 10 & 213 & 1.01\% & 1.23\% & 0.59\% & 1.01\% \\
CoNaLa & 12 & 220 & 314 & 190 & 2749 & 2 & 2 & 2 & 2 & 0.97\% & 1.11\% & 0.68\% & 0.95\% \\
Headlines & 13 & 734 & 1098 & 659 & 9900 & 4 & 6 & 3 & 56 & 0.57\% & 0.72\% & 0.44\% & 0.57\% \\
Overruling & 12 & 193 & 264 & 165 & 2372 & 2 & 3 & 1 & 23 & 0.99\% & 1.23\% & 0.77\% & 0.99\% \\
\midrule[1pt]
\end{tabular}%
}
\vspace{-6pt}
\end{table}
\vspace{-6pt}
\end{center}
}

By default, the framework is to prioritize the optimization of the performance-to-cost ratio. However, users can implement alternative allocation strategies. We introduce two alternative scheduling rules, including the Performance-First (Max-Perf) rule and the Cost-Optimized (Min-Cost) rule. The comparison of SLS adopting different scheduling rules is reported in Table~\ref{tb_res_rule}. 

The Max-Perf rule selects the LLM with the highest predicted performance, prioritizing outcome quality regardless of cost. While this rule results in a modest performance improvement over the default setting, it leads to significantly higher costs. For example, on the Proxifier dataset, SLS improves performance by 2.09\%, increasing from 0.7545 to 0.7703. However, this gain comes with a substantial cost increase, from \$9.49 to \$42.58, and an extended makespan, increasing from 91 minutes to 241 minutes. 
The Min-Cost rule selects the least expensive option among the predicted feasible candidates, defined as those with a predicted probability of success exceeding 0.5. This rule prioritizes cost efficiency by avoiding unnecessary expenditure on queries unlikely to succeed. When all candidate LLMs are predicted to have a low likelihood of success, the scheduler conducts not to submit the query to any model, thereby preventing wasteful invocations. 
For example, on the CoNaLa dataset, the cost is reduced from \$0.81 to \$0.26, as the scheduler predicted that many queries could not be solved by any available LLM, and therefore did not submit them.

 We also evaluate the choice of classifier on the overall performance of SLS.  We assess SLS using Stochastic Gradient Descent (SGD)~\cite{zhang2004solving}, which is an efficient optimization algorithm for training linear classifiers and regressors. \rv{The detailed results can be found on our project Webpage.} The results show that while XGBoost has demonstrated superior performance in most cases, the SLS framework offers significant flexibility, enabling users to tailor the prediction model by selecting alternative classifiers.
 \begin{center}
\begin{table}[ht]
\caption{The performance of SLS with different scheduling rules}
 \vspace{-6pt}
\label{tb_res_rule}
\resizebox{1\columnwidth}{!}{%
\begin{tabular}{lllllllllll}
\midrule[1pt]
\multicolumn{1}{c}{\multirow{2}{*}{Dataset}} & \multicolumn{1}{c}{\multirow{2}{*}{Rule}} & \multicolumn{3}{c}{Performance} & \multicolumn{3}{c}{Total Cost} & \multicolumn{3}{c}{Makespan} \\
\multicolumn{1}{c}{} & \multicolumn{1}{c}{} & \multicolumn{1}{c}{Avg.} & \multicolumn{1}{c}{Max} & \multicolumn{1}{c}{Min} & \multicolumn{1}{c}{Avg} & \multicolumn{1}{c}{Max} & \multicolumn{1}{c}{Min} & \multicolumn{1}{c}{Avg} & \multicolumn{1}{c}{Max} & \multicolumn{1}{c}{Min} \\ 
\midrule[1pt]
Apache & Default & 0.9968 & 0.9981 & 0.9949 & \textbf{2.42} & \textbf{2.48} & \textbf{2.40} & \textbf{34} & 85 & 26 \\
 & Min-Cost & 0.9974 & 0.9982 & 0.9961 & 2.44 & 2.50 & 2.42 & 47 & 59 & 38 \\
 & Max-Perf & \textbf{0.9984} & \textbf{0.9989} & \textbf{0.9976} & 4.20 & 5.29 & 3.59 & 36 & \textbf{49} & \textbf{29} \\ \hline
\multirow{3}{*}{Linux} & Default & 0.9183 & 0.9303 & \textbf{0.9051} & 1.51 & 1.59 & 1.44 & 41 & 47 & 33 \\
 & Min-Cost & 0.9112 & 0.9239 & 0.8926 & \textbf{1.45} & \textbf{1.56} & \textbf{1.37} & \textbf{30} & \textbf{37} & \textbf{26} \\
 & Max-Perf & \textbf{0.9222} & \textbf{0.9318} & 0.9027 & 2.40 & 2.83 & 1.95 & 33 & 41 & 26 \\ \hline
\multirow{3}{*}{Proxifier} & Default & 0.7545 & 0.7916 & 0.6959 & 9.49 & 13.70 & 7.28 & 91 & 138 & 71 \\
 & Min-Cost & 0.6921 & 0.7194 & 0.6340 & \multicolumn{1}{l}{\textbf{6.30}} & \multicolumn{1}{l}{\textbf{10.25}} & \multicolumn{1}{l}{\textbf{3.44}} & \multicolumn{1}{l}{\textbf{42}} & \multicolumn{1}{l}{\textbf{67}} & \multicolumn{1}{l}{\textbf{24}} \\
 & Max-Perf & \textbf{0.7703} & \textbf{0.8090} & \textbf{0.7279} & \multicolumn{1}{l}{42.58} & \multicolumn{1}{l}{79.58} & \multicolumn{1}{l}{16.92} & \multicolumn{1}{l}{241} & \multicolumn{1}{l}{482} & \multicolumn{1}{l}{86} \\ \hline
\multirow{3}{*}{CoNaLa} & Default & 0.1587 & 0.1597 & 0.1574 & \multicolumn{1}{l}{0.81} & \multicolumn{1}{l}{0.85} & \multicolumn{1}{l}{0.78} & \multicolumn{1}{l}{21} & \multicolumn{1}{l}{22} & \multicolumn{1}{l}{20} \\
 & Min-Cost & 0.1284 & 0.1294 & 0.1276 & \textbf{0.26} & \textbf{0.30} & \textbf{0.24} & \textbf{5} & \textbf{6} & \textbf{4} \\
 & Max-Perf & \textbf{0.2000} & \textbf{0.2010} & \textbf{0.1992} & 1.31 & 1.35 & 1.28 & 14 & 15 & 14 \\ \hline
\multirow{3}{*}{Headlines} & Default & \multicolumn{1}{l}{0.7890} & \multicolumn{1}{l}{0.8000} & \multicolumn{1}{l}{0.7709} & \multicolumn{1}{l}{\textbf{0.50}} & \multicolumn{1}{l}{\textbf{0.64}} & \multicolumn{1}{l}{\textbf{0.36}} & \multicolumn{1}{l}{50} & \multicolumn{1}{l}{84} & \multicolumn{1}{l}{33} \\
 & Min-Cost & 0.7756 & 0.7891 & 0.7645 & 0.60 & 0.86 & 0.37 & \textbf{48} & \textbf{60} & \textbf{41} \\
 & Max-Perf & \textbf{0.8109} & \textbf{0.8258} & \textbf{0.7996} & 3.61 & 5.58 & 1.25 & 52 & 80 & 36 \\ \hline
\multirow{3}{*}{Overruling} & Default & 0.9584 & 0.9760 & 0.9389 & 0.89 & 1.42 & 0.42 & 26 & 39 & 20 \\
 & Min-Cost & 0.9460 & 0.9469 & \textbf{0.9414} & \textbf{0.3} & \textbf{0.32} & \textbf{0.27} & \textbf{18} & \textbf{23} & \textbf{16} \\
 & Max-Perf & \textbf{0.9678} & \textbf{0.9798} & 0.9406 & 1.53 & 1.72 & 1.34 & 32 & 37 & 31 \\ 
 \midrule[1pt]
\end{tabular}%
}
\vspace{-12pt}
\end{table}

\end{center}

\subsection{Applicability Across Deployment Settings}
The current implementation of SLS assumes API-based access to LLMs, where cost is explicitly defined based on the number of input and output tokens. This setting is representative of many commercial LLM services and enables clear cost-performance trade-offs during scheduling. In locally hosted environments, while direct monetary cost may no longer apply, other resource-related objectives such as makespan, GPU usage, memory consumption, or energy efficiency remain critical. The scheduler component of SLS can be adapted to optimize these objectives by replacing the objective function in the scheduling module. Therefore, although the current cost definition is API-specific, the overall framework remains applicable and valuable for local deployment settings by incorporating resource-aware cost estimations. Future work may further explore hybrid environments where both API-based and local models coexist, requiring more complex scheduling decisions across heterogeneous execution platforms.

\subsection{Threats to Validity}

\textbf{Prompt Design.}
The design of input prompts significantly influences the performance and cost of invoking LLMs. In this work, we use the same prompt for all candidate LLMs for each job type. As a direction for future research, we propose incorporating a dedicated Prompt Manager to generate prompts with varying complexity levels and structures for training and evaluation. This enhancement would enable the framework to handle a broader range of scenarios and improve its robustness in different application contexts.

\noindent\textbf{LLM-Specific Constraints.} The overall performance of the SLS framework is inherently dependent on the candidate LLMs available for selection. If the LLMs perform poorly on certain job types, the effectiveness of the framework may be compromised. 

\rv{\noindent\textbf{Dependence on Labeled Data.} This dependence on labeled data limits the immediate applicability of SLS in fully automated or labeling-free environments. Future work could explore integrating label-free techniques such as self-speculative feedback~\cite{guha2024smoothie} or correctness prediction based on benchmark evaluations~\cite{shnitzer2023large} into SLS to reduce reliance on labeled data and enhance scalability.}

\vspace{-6pt}
\section{Conclusion}
The impressive capabilities of LLMs highlight their potential in various intelligent software engineering tasks. However, the high operational costs associated with deploying LLMs remain a crucial barrier.  
This paper presents the SLS framework, which assigns each query to the most suitable LLM and reuses previously computed results through adaptive caching to enhance performance. Experimental results show that SLS improves performance, reduces invocation costs, and minimizes response times compared to the baseline methods.

\balance
\bibliographystyle{ACM-Reference-Format}
\bibliography{library}


\begin{thebibliography}{45}


\ifx \showCODEN    \undefined \def \showCODEN     #1{\unskip}     \fi
\ifx \showDOI      \undefined \def \showDOI       #1{#1}\fi
\ifx \showISBNx    \undefined \def \showISBNx     #1{\unskip}     \fi
\ifx \showISBNxiii \undefined \def \showISBNxiii  #1{\unskip}     \fi
\ifx \showISSN     \undefined \def \showISSN      #1{\unskip}     \fi
\ifx \showLCCN     \undefined \def \showLCCN      #1{\unskip}     \fi
\ifx \shownote     \undefined \def \shownote      #1{#1}          \fi
\ifx \showarticletitle \undefined \def \showarticletitle #1{#1}   \fi
\ifx \showURL      \undefined \def \showURL       {\relax}        \fi
\providecommand\bibfield[2]{#2}
\providecommand\bibinfo[2]{#2}
\providecommand\natexlab[1]{#1}
\providecommand\showeprint[2][]{arXiv:#2}

\bibitem[ant(2024)]%
        {anthropic_pricing}
 \bibinfo{year}{2024}\natexlab{}.
\newblock \bibinfo{booktitle}{\emph{Claude Pricing}}.
\newblock
\urldef\tempurl%
\url{https://www.anthropic.com/pricing}
\showURL{%
Retrieved August 2, 2024 from \tempurl}


\bibitem[Baek et~al\mbox{.}(2024)]%
        {baek2024knowledge}
\bibfield{author}{\bibinfo{person}{Jinheon Baek}, \bibinfo{person}{Nirupama Chandrasekaran}, \bibinfo{person}{Silviu Cucerzan}, \bibinfo{person}{Allen Herring}, {and} \bibinfo{person}{Sujay~Kumar Jauhar}.} \bibinfo{year}{2024}\natexlab{}.
\newblock \showarticletitle{Knowledge-augmented large language models for personalized contextual query suggestion}. In \bibinfo{booktitle}{\emph{Proceedings of the ACM on Web Conference 2024}}. \bibinfo{pages}{3355--3366}.
\newblock


\bibitem[Bang(2023)]%
        {bang2023gptcache}
\bibfield{author}{\bibinfo{person}{Fu Bang}.} \bibinfo{year}{2023}\natexlab{}.
\newblock \showarticletitle{GPTCache: An open-source semantic cache for LLM applications enabling faster answers and cost savings}. In \bibinfo{booktitle}{\emph{Proceedings of the 3rd Workshop for Natural Language Processing Open Source Software (NLP-OSS 2023)}}. \bibinfo{pages}{212--218}.
\newblock


\bibitem[Chen et~al\mbox{.}(2024)]%
        {chen2024frugalgpt}
\bibfield{author}{\bibinfo{person}{Lingjiao Chen}, \bibinfo{person}{Matei Zaharia}, {and} \bibinfo{person}{James Zou}.} \bibinfo{year}{2024}\natexlab{}.
\newblock \showarticletitle{FrugalGPT: How to Use Large Language Models While Reducing Cost and Improving Performance}.
\newblock \bibinfo{journal}{\emph{Transactions on Machine Learning Research}} (\bibinfo{year}{2024}).
\newblock


\bibitem[Fakhoury et~al\mbox{.}(2024)]%
        {fakhoury2024llm}
\bibfield{author}{\bibinfo{person}{Sarah Fakhoury}, \bibinfo{person}{Aaditya Naik}, \bibinfo{person}{Georgios Sakkas}, \bibinfo{person}{Saikat Chakraborty}, {and} \bibinfo{person}{Shuvendu~K Lahiri}.} \bibinfo{year}{2024}\natexlab{}.
\newblock \showarticletitle{Llm-based test-driven interactive code generation: User study and empirical evaluation}.
\newblock \bibinfo{journal}{\emph{IEEE Transactions on Software Engineering}} (\bibinfo{year}{2024}).
\newblock


\bibitem[Feng et~al\mbox{.}(2020)]%
        {feng-etal-2020-codebert}
\bibfield{author}{\bibinfo{person}{Zhangyin Feng}, \bibinfo{person}{Daya Guo}, \bibinfo{person}{Duyu Tang}, \bibinfo{person}{Nan Duan}, \bibinfo{person}{Xiaocheng Feng}, \bibinfo{person}{Ming Gong}, \bibinfo{person}{Linjun Shou}, \bibinfo{person}{Bing Qin}, \bibinfo{person}{Ting Liu}, \bibinfo{person}{Daxin Jiang}, {and} \bibinfo{person}{Ming Zhou}.} \bibinfo{year}{2020}\natexlab{}.
\newblock \showarticletitle{{C}ode{BERT}: A Pre-Trained Model for Programming and Natural Languages}. In \bibinfo{booktitle}{\emph{Findings of the Association for Computational Linguistics: EMNLP 2020}}, \bibfield{editor}{\bibinfo{person}{Trevor Cohn}, \bibinfo{person}{Yulan He}, {and} \bibinfo{person}{Yang Liu}} (Eds.). \bibinfo{publisher}{Association for Computational Linguistics}, \bibinfo{address}{Online}, \bibinfo{pages}{1536--1547}.
\newblock
\urldef\tempurl%
\url{https://doi.org/10.18653/v1/2020.findings-emnlp.139}
\showDOI{\tempurl}


\bibitem[Fu et~al\mbox{.}(2024)]%
        {fu2024missconf}
\bibfield{author}{\bibinfo{person}{Ying Fu}, \bibinfo{person}{Teng Wang}, \bibinfo{person}{Shanshan Li}, \bibinfo{person}{Jinyan Ding}, \bibinfo{person}{Shulin Zhou}, \bibinfo{person}{Zhouyang Jia}, \bibinfo{person}{Wang Li}, \bibinfo{person}{Yu Jiang}, {and} \bibinfo{person}{Xiangke Liao}.} \bibinfo{year}{2024}\natexlab{}.
\newblock \showarticletitle{MissConf: LLM-Enhanced Reproduction of Configuration-Triggered Bugs}. In \bibinfo{booktitle}{\emph{Proceedings of the 2024 IEEE/ACM 46th International Conference on Software Engineering: Companion Proceedings}}. \bibinfo{pages}{484--495}.
\newblock


\bibitem[Gill et~al\mbox{.}(2024)]%
        {gill2024privacy}
\bibfield{author}{\bibinfo{person}{Waris Gill}, \bibinfo{person}{Mohamed Elidrisi}, \bibinfo{person}{Pallavi Kalapatapu}, \bibinfo{person}{Ali Anwar}, {and} \bibinfo{person}{Muhammad~Ali Gulzar}.} \bibinfo{year}{2024}\natexlab{}.
\newblock \showarticletitle{Privacy-Aware Semantic Cache for Large Language Models}.
\newblock \bibinfo{journal}{\emph{arXiv preprint arXiv:2403.02694}} (\bibinfo{year}{2024}).
\newblock


\bibitem[Glomb et~al\mbox{.}(2022)]%
        {glomb2022rolling}
\bibfield{author}{\bibinfo{person}{Lukas Glomb}, \bibinfo{person}{Frauke Liers}, {and} \bibinfo{person}{Florian R{\"o}sel}.} \bibinfo{year}{2022}\natexlab{}.
\newblock \showarticletitle{A rolling-horizon approach for multi-period optimization}.
\newblock \bibinfo{journal}{\emph{European Journal of Operational Research}} \bibinfo{volume}{300}, \bibinfo{number}{1} (\bibinfo{year}{2022}), \bibinfo{pages}{189--206}.
\newblock


\bibitem[Guha et~al\mbox{.}(2024)]%
        {guha2024smoothie}
\bibfield{author}{\bibinfo{person}{Neel Guha}, \bibinfo{person}{Mayee Chen}, \bibinfo{person}{Trevor Chow}, \bibinfo{person}{Ishan Khare}, {and} \bibinfo{person}{Christopher Re}.} \bibinfo{year}{2024}\natexlab{}.
\newblock \showarticletitle{Smoothie: Label free language model routing}.
\newblock \bibinfo{journal}{\emph{Advances in Neural Information Processing Systems}}  \bibinfo{volume}{37} (\bibinfo{year}{2024}), \bibinfo{pages}{127645--127672}.
\newblock


\bibitem[Guo et~al\mbox{.}(2024)]%
        {guo2024outpatient}
\bibfield{author}{\bibinfo{person}{Hainan Guo}, \bibinfo{person}{Yue Xie}, \bibinfo{person}{Bowen Jiang}, {and} \bibinfo{person}{Jiafu Tang}.} \bibinfo{year}{2024}\natexlab{}.
\newblock \showarticletitle{When outpatient appointment meets online consultation: A joint scheduling optimization framework}.
\newblock \bibinfo{journal}{\emph{Omega}}  \bibinfo{volume}{127} (\bibinfo{year}{2024}), \bibinfo{pages}{103101}.
\newblock


\bibitem[Hou et~al\mbox{.}(2023)]%
        {hou2023large}
\bibfield{author}{\bibinfo{person}{Xinyi Hou}, \bibinfo{person}{Yanjie Zhao}, \bibinfo{person}{Yue Liu}, \bibinfo{person}{Zhou Yang}, \bibinfo{person}{Kailong Wang}, \bibinfo{person}{Li Li}, \bibinfo{person}{Xiapu Luo}, \bibinfo{person}{David Lo}, \bibinfo{person}{John Grundy}, {and} \bibinfo{person}{Haoyu Wang}.} \bibinfo{year}{2023}\natexlab{}.
\newblock \showarticletitle{Large language models for software engineering: A systematic literature review}.
\newblock \bibinfo{journal}{\emph{ACM Transactions on Software Engineering and Methodology}} (\bibinfo{year}{2023}).
\newblock


\bibitem[Jiang et~al\mbox{.}(2024)]%
        {jiang2024survey}
\bibfield{author}{\bibinfo{person}{Juyong Jiang}, \bibinfo{person}{Fan Wang}, \bibinfo{person}{Jiasi Shen}, \bibinfo{person}{Sungju Kim}, {and} \bibinfo{person}{Sunghun Kim}.} \bibinfo{year}{2024}\natexlab{}.
\newblock \showarticletitle{A survey on large language models for code generation}.
\newblock \bibinfo{journal}{\emph{arXiv preprint arXiv:2406.00515}} (\bibinfo{year}{2024}).
\newblock


\bibitem[Kang et~al\mbox{.}(2023)]%
        {kang2023large}
\bibfield{author}{\bibinfo{person}{Sungmin Kang}, \bibinfo{person}{Juyeon Yoon}, {and} \bibinfo{person}{Shin Yoo}.} \bibinfo{year}{2023}\natexlab{}.
\newblock \showarticletitle{Large language models are few-shot testers: Exploring llm-based general bug reproduction}. In \bibinfo{booktitle}{\emph{2023 IEEE/ACM 45th International Conference on Software Engineering (ICSE)}}. IEEE, \bibinfo{pages}{2312--2323}.
\newblock


\bibitem[Karmakar and Robbes(2021)]%
        {karmakar2021pre}
\bibfield{author}{\bibinfo{person}{Anjan Karmakar} {and} \bibinfo{person}{Romain Robbes}.} \bibinfo{year}{2021}\natexlab{}.
\newblock \showarticletitle{What do pre-trained code models know about code?}. In \bibinfo{booktitle}{\emph{Proceedings of the 36th IEEE/ACM International Conference on Automated Software Engineering (ASE)}}. IEEE, \bibinfo{pages}{1332--1336}.
\newblock


\bibitem[Liang et~al\mbox{.}(2023)]%
        {liang2023poisson}
\bibfield{author}{\bibinfo{person}{Jian Liang}, \bibinfo{person}{Jintao Ke}, \bibinfo{person}{Hai Wang}, \bibinfo{person}{Hongbo Ye}, {and} \bibinfo{person}{Jinjun Tang}.} \bibinfo{year}{2023}\natexlab{}.
\newblock \showarticletitle{A Poisson-based distribution learning framework for short-term prediction of food delivery demand ranges}.
\newblock \bibinfo{journal}{\emph{IEEE Transactions on Intelligent Transportation Systems}} (\bibinfo{year}{2023}).
\newblock


\bibitem[Liu et~al\mbox{.}(2024)]%
        {liu2024optllm}
\bibfield{author}{\bibinfo{person}{Yueyue Liu}, \bibinfo{person}{Hongyu Zhang}, \bibinfo{person}{Yuantian Miao}, \bibinfo{person}{Van-Hoang Le}, {and} \bibinfo{person}{Zhiqiang Li}.} \bibinfo{year}{2024}\natexlab{}.
\newblock \showarticletitle{OptLLM: Optimal Assignment of Queries to Large Language Models}. In \bibinfo{booktitle}{\emph{2024 IEEE International Conference on Web Services (ICWS)}}. IEEE, \bibinfo{pages}{788--798}.
\newblock


\bibitem[Ma et~al\mbox{.}(2024)]%
        {ma2024llmparser}
\bibfield{author}{\bibinfo{person}{Zeyang Ma}, \bibinfo{person}{An~Ran Chen}, \bibinfo{person}{Dong~Jae Kim}, \bibinfo{person}{Tse-Hsun Chen}, {and} \bibinfo{person}{Shaowei Wang}.} \bibinfo{year}{2024}\natexlab{}.
\newblock \showarticletitle{Llmparser: An exploratory study on using large language models for log parsing}. In \bibinfo{booktitle}{\emph{Proceedings of the IEEE/ACM 46th International Conference on Software Engineering (ICSE)}}. \bibinfo{pages}{1--13}.
\newblock


\bibitem[Mashhadi and Hemmati(2021)]%
        {mashhadi2021applying}
\bibfield{author}{\bibinfo{person}{Ehsan Mashhadi} {and} \bibinfo{person}{Hadi Hemmati}.} \bibinfo{year}{2021}\natexlab{}.
\newblock \showarticletitle{Applying codebert for automated program repair of java simple bugs}. In \bibinfo{booktitle}{\emph{2021 IEEE/ACM 18th International Conference on Mining Software Repositories (MSR)}}. IEEE, \bibinfo{pages}{505--509}.
\newblock


\bibitem[Mu et~al\mbox{.}(2024)]%
        {mu2024clarifygpt}
\bibfield{author}{\bibinfo{person}{Fangwen Mu}, \bibinfo{person}{Lin Shi}, \bibinfo{person}{Song Wang}, \bibinfo{person}{Zhuohao Yu}, \bibinfo{person}{Binquan Zhang}, \bibinfo{person}{ChenXue Wang}, \bibinfo{person}{Shichao Liu}, {and} \bibinfo{person}{Qing Wang}.} \bibinfo{year}{2024}\natexlab{}.
\newblock \showarticletitle{Clarifygpt: A framework for enhancing llm-based code generation via requirements clarification}.
\newblock \bibinfo{journal}{\emph{Proceedings of the ACM on Software Engineering}} \bibinfo{volume}{1}, \bibinfo{number}{FSE} (\bibinfo{year}{2024}), \bibinfo{pages}{2332--2354}.
\newblock


\bibitem[Nam et~al\mbox{.}(2024)]%
        {nam2024using}
\bibfield{author}{\bibinfo{person}{Daye Nam}, \bibinfo{person}{Andrew Macvean}, \bibinfo{person}{Vincent Hellendoorn}, \bibinfo{person}{Bogdan Vasilescu}, {and} \bibinfo{person}{Brad Myers}.} \bibinfo{year}{2024}\natexlab{}.
\newblock \showarticletitle{Using an llm to help with code understanding}. In \bibinfo{booktitle}{\emph{Proceedings of the IEEE/ACM 46th International Conference on Software Engineering (ICSE)}}. \bibinfo{pages}{1--13}.
\newblock


\bibitem[Pan and Yang(2005)]%
        {pan2005fifo}
\bibfield{author}{\bibinfo{person}{Deng Pan} {and} \bibinfo{person}{Yuanyuan Yang}.} \bibinfo{year}{2005}\natexlab{}.
\newblock \showarticletitle{FIFO-based multicast scheduling algorithm for virtual output queued packet switches}.
\newblock \bibinfo{journal}{\emph{IEEE Trans. Comput.}} \bibinfo{volume}{54}, \bibinfo{number}{10} (\bibinfo{year}{2005}), \bibinfo{pages}{1283--1297}.
\newblock


\bibitem[Papineni et~al\mbox{.}(2002)]%
        {papineni2002bleu}
\bibfield{author}{\bibinfo{person}{Kishore Papineni}, \bibinfo{person}{Salim Roukos}, \bibinfo{person}{Todd Ward}, {and} \bibinfo{person}{Wei-Jing Zhu}.} \bibinfo{year}{2002}\natexlab{}.
\newblock \showarticletitle{Bleu: a method for automatic evaluation of machine translation}. In \bibinfo{booktitle}{\emph{Proceedings of the 40th annual meeting of the Association for Computational Linguistics}}. \bibinfo{pages}{311--318}.
\newblock


\bibitem[Qu et~al\mbox{.}(2024)]%
        {qu2024cloud}
\bibfield{author}{\bibinfo{person}{Zihao Qu}, \bibinfo{person}{Milind Dawande}, {and} \bibinfo{person}{Ganesh Janakiraman}.} \bibinfo{year}{2024}\natexlab{}.
\newblock \showarticletitle{Cloud Cost Optimization: Model, Bounds, and Asymptotics}.
\newblock \bibinfo{journal}{\emph{Operations Research}} \bibinfo{volume}{72}, \bibinfo{number}{1} (\bibinfo{year}{2024}), \bibinfo{pages}{132--150}.
\newblock


\bibitem[Reimers and Gurevych(2019)]%
        {Reimers2019SentenceBERTSE}
\bibfield{author}{\bibinfo{person}{Nils Reimers} {and} \bibinfo{person}{Iryna Gurevych}.} \bibinfo{year}{2019}\natexlab{}.
\newblock \showarticletitle{Sentence-BERT: Sentence Embeddings using Siamese BERT-Networks}. In \bibinfo{booktitle}{\emph{Conference on Empirical Methods in Natural Language Processing}}.
\newblock
\urldef\tempurl%
\url{https://api.semanticscholar.org/CorpusID:201646309}
\showURL{%
\tempurl}


\bibitem[Roy et~al\mbox{.}(2024)]%
        {roy2024exploring}
\bibfield{author}{\bibinfo{person}{Devjeet Roy}, \bibinfo{person}{Xuchao Zhang}, \bibinfo{person}{Rashi Bhave}, \bibinfo{person}{Chetan Bansal}, \bibinfo{person}{Pedro Las-Casas}, \bibinfo{person}{Rodrigo Fonseca}, {and} \bibinfo{person}{Saravan Rajmohan}.} \bibinfo{year}{2024}\natexlab{}.
\newblock \showarticletitle{Exploring llm-based agents for root cause analysis}. In \bibinfo{booktitle}{\emph{Companion Proceedings of the 32nd ACM International Conference on the Foundations of Software Engineering}}. \bibinfo{pages}{208--219}.
\newblock


\bibitem[{\v{S}}akota et~al\mbox{.}(2024)]%
        {vsakota2024fly}
\bibfield{author}{\bibinfo{person}{Marija {\v{S}}akota}, \bibinfo{person}{Maxime Peyrard}, {and} \bibinfo{person}{Robert West}.} \bibinfo{year}{2024}\natexlab{}.
\newblock \showarticletitle{Fly-swat or cannon? cost-effective language model choice via meta-modeling}. In \bibinfo{booktitle}{\emph{Proceedings of the 17th ACM International Conference on Web Search and Data Mining}}. \bibinfo{pages}{606--615}.
\newblock


\bibitem[Shnitzer et~al\mbox{.}(2023)]%
        {shnitzer2023large}
\bibfield{author}{\bibinfo{person}{Tal Shnitzer}, \bibinfo{person}{Anthony Ou}, \bibinfo{person}{M{\'\i}rian Silva}, \bibinfo{person}{Kate Soule}, \bibinfo{person}{Yuekai Sun}, \bibinfo{person}{Justin Solomon}, \bibinfo{person}{Neil Thompson}, {and} \bibinfo{person}{Mikhail Yurochkin}.} \bibinfo{year}{2023}\natexlab{}.
\newblock \showarticletitle{Large language model routing with benchmark datasets}.
\newblock \bibinfo{journal}{\emph{arXiv preprint arXiv:2309.15789}} (\bibinfo{year}{2023}).
\newblock


\bibitem[Sinha and Khandait(2021)]%
        {sinha2021impact}
\bibfield{author}{\bibinfo{person}{Ankur Sinha} {and} \bibinfo{person}{Tanmay Khandait}.} \bibinfo{year}{2021}\natexlab{}.
\newblock \showarticletitle{Impact of news on the commodity market: Dataset and results}. In \bibinfo{booktitle}{\emph{Advances in Information and Communication: Proceedings of the 2021 Future of Information and Communication Conference (FICC), Volume 2}}. Springer, \bibinfo{pages}{589--601}.
\newblock


\bibitem[Su et~al\mbox{.}(2023)]%
        {su2023multi}
\bibfield{author}{\bibinfo{person}{Zhixiang Su}, \bibinfo{person}{Di Wang}, \bibinfo{person}{Chunyan Miao}, {and} \bibinfo{person}{Lizhen Cui}.} \bibinfo{year}{2023}\natexlab{}.
\newblock \showarticletitle{Multi-aspect explainable inductive relation prediction by sentence transformer}. In \bibinfo{booktitle}{\emph{Proceedings of the AAAI Conference on Artificial Intelligence}}, Vol.~\bibinfo{volume}{37}. \bibinfo{pages}{6533--6540}.
\newblock


\bibitem[Sun et~al\mbox{.}(2024)]%
        {sun2024consistency}
\bibfield{author}{\bibinfo{person}{Qi Sun}, \bibinfo{person}{Kun Huang}, \bibinfo{person}{Xiaocui Yang}, \bibinfo{person}{Rong Tong}, \bibinfo{person}{Kun Zhang}, {and} \bibinfo{person}{Soujanya Poria}.} \bibinfo{year}{2024}\natexlab{}.
\newblock \showarticletitle{Consistency guided knowledge retrieval and denoising in llms for zero-shot document-level relation triplet extraction}. In \bibinfo{booktitle}{\emph{Proceedings of the ACM on Web Conference 2024}}. \bibinfo{pages}{4407--4416}.
\newblock


\bibitem[Wang et~al\mbox{.}(2024)]%
        {wang2024survey}
\bibfield{author}{\bibinfo{person}{Can Wang}, \bibinfo{person}{Bolin Zhang}, \bibinfo{person}{Dianbo Sui}, \bibinfo{person}{Zhiying Tum}, \bibinfo{person}{Xiaoyu Liu}, {and} \bibinfo{person}{Jiabao Kang}.} \bibinfo{year}{2024}\natexlab{}.
\newblock \showarticletitle{A Survey on Effective Invocation Methods of Massive LLM Services}.
\newblock \bibinfo{journal}{\emph{arXiv preprint arXiv:2402.03408}} (\bibinfo{year}{2024}).
\newblock


\bibitem[Wu et~al\mbox{.}(2024)]%
        {wu2024unify}
\bibfield{author}{\bibinfo{person}{Songhao Wu}, \bibinfo{person}{Quan Tu}, \bibinfo{person}{Hong Liu}, \bibinfo{person}{Jia Xu}, \bibinfo{person}{Zhongyi Liu}, \bibinfo{person}{Guannan Zhang}, \bibinfo{person}{Ran Wang}, \bibinfo{person}{Xiuying Chen}, {and} \bibinfo{person}{Rui Yan}.} \bibinfo{year}{2024}\natexlab{}.
\newblock \showarticletitle{Unify Graph Learning with Text: Unleashing LLM Potentials for Session Search}. In \bibinfo{booktitle}{\emph{Proceedings of the ACM on Web Conference 2024}}. \bibinfo{pages}{1509--1518}.
\newblock


\bibitem[Xia and Zhang(2024)]%
        {xia2024automated}
\bibfield{author}{\bibinfo{person}{Chunqiu~Steven Xia} {and} \bibinfo{person}{Lingming Zhang}.} \bibinfo{year}{2024}\natexlab{}.
\newblock \showarticletitle{Automated program repair via conversation: Fixing 162 out of 337 bugs for \$0.42 each using ChatGPT}. In \bibinfo{booktitle}{\emph{Proceedings of the 33rd ACM SIGSOFT International Symposium on Software Testing and Analysis}}. \bibinfo{pages}{819--831}.
\newblock


\bibitem[Xia et~al\mbox{.}(2024)]%
        {xia2024llm}
\bibfield{author}{\bibinfo{person}{Yu Xia}, \bibinfo{person}{Fang Kong}, \bibinfo{person}{Tong Yu}, \bibinfo{person}{Liya Guo}, \bibinfo{person}{Ryan~A Rossi}, \bibinfo{person}{Sungchul Kim}, {and} \bibinfo{person}{Shuai Li}.} \bibinfo{year}{2024}\natexlab{}.
\newblock \showarticletitle{Which LLM to Play? Convergence-Aware Online Model Selection with Time-Increasing Bandits}. In \bibinfo{booktitle}{\emph{Proceedings of the ACM on Web Conference 2024}}. \bibinfo{pages}{4059--4070}.
\newblock


\bibitem[Xiao et~al\mbox{.}(2024)]%
        {xiao2024free}
\bibfield{author}{\bibinfo{person}{Yi Xiao}, \bibinfo{person}{Van-Hoang Le}, {and} \bibinfo{person}{Hongyu Zhang}.} \bibinfo{year}{2024}\natexlab{}.
\newblock \showarticletitle{Free: Towards More Practical Log Parsing with Large Language Models}. In \bibinfo{booktitle}{\emph{Proceedings of the 39th IEEE/ACM International Conference on Automated Software Engineering (ASE)}}. \bibinfo{pages}{153--165}.
\newblock


\bibitem[Xu et~al\mbox{.}(2024)]%
        {xu2024divlog}
\bibfield{author}{\bibinfo{person}{Junjielong Xu}, \bibinfo{person}{Ruichun Yang}, \bibinfo{person}{Yintong Huo}, \bibinfo{person}{Chengyu Zhang}, {and} \bibinfo{person}{Pinjia He}.} \bibinfo{year}{2024}\natexlab{}.
\newblock \showarticletitle{Divlog: Log parsing with prompt enhanced in-context learning}. In \bibinfo{booktitle}{\emph{Proceedings of the IEEE/ACM 46th International Conference on Software Engineering (ICSE)}}. \bibinfo{pages}{1--12}.
\newblock


\bibitem[Yadav et~al\mbox{.}(2024)]%
        {yadav2024pag}
\bibfield{author}{\bibinfo{person}{Vikas Yadav}, \bibinfo{person}{Zheng Tang}, {and} \bibinfo{person}{Vijay Srinivasan}.} \bibinfo{year}{2024}\natexlab{}.
\newblock \showarticletitle{PAG-LLM: Paraphrase and Aggregate with Large Language Models for Minimizing Intent Classification Errors}. In \bibinfo{booktitle}{\emph{Proceedings of the 47th International ACM SIGIR Conference on Research and Development in Information Retrieval}}. \bibinfo{pages}{2569--2573}.
\newblock


\bibitem[Yin et~al\mbox{.}(2018)]%
        {yin2018learning}
\bibfield{author}{\bibinfo{person}{Pengcheng Yin}, \bibinfo{person}{Bowen Deng}, \bibinfo{person}{Edgar Chen}, \bibinfo{person}{Bogdan Vasilescu}, {and} \bibinfo{person}{Graham Neubig}.} \bibinfo{year}{2018}\natexlab{}.
\newblock \showarticletitle{Learning to mine aligned code and natural language pairs from stack overflow}. In \bibinfo{booktitle}{\emph{Proceedings of the 15th international conference on mining software repositories}}. \bibinfo{pages}{476--486}.
\newblock


\bibitem[Zhang(2004)]%
        {zhang2004solving}
\bibfield{author}{\bibinfo{person}{Tong Zhang}.} \bibinfo{year}{2004}\natexlab{}.
\newblock \showarticletitle{Solving large scale linear prediction problems using stochastic gradient descent algorithms}. In \bibinfo{booktitle}{\emph{Proceedings of the twenty-first international conference on Machine learning}}. \bibinfo{pages}{116}.
\newblock


\bibitem[Zhang et~al\mbox{.}({[n.\,d.]})]%
        {zhangcam}
\bibfield{author}{\bibinfo{person}{Yuxin Zhang}, \bibinfo{person}{Yuxuan Du}, \bibinfo{person}{Gen Luo}, \bibinfo{person}{Yunshan Zhong}, \bibinfo{person}{Zhenyu Zhang}, \bibinfo{person}{Shiwei Liu}, {and} \bibinfo{person}{Rongrong Ji}.} \bibinfo{year}{[n.\,d.]}\natexlab{}.
\newblock \showarticletitle{CaM: Cache Merging for Memory-efficient LLMs Inference}. In \bibinfo{booktitle}{\emph{Forty-first International Conference on Machine Learning (ICML2024)}}.
\newblock


\bibitem[Zheng et~al\mbox{.}(2021)]%
        {zheng2021does}
\bibfield{author}{\bibinfo{person}{Lucia Zheng}, \bibinfo{person}{Neel Guha}, \bibinfo{person}{Brandon~R Anderson}, \bibinfo{person}{Peter Henderson}, {and} \bibinfo{person}{Daniel~E Ho}.} \bibinfo{year}{2021}\natexlab{}.
\newblock \showarticletitle{When does pretraining help? assessing self-supervised learning for law and the casehold dataset of 53,000+ legal holdings}. In \bibinfo{booktitle}{\emph{Proceedings of the 18th International Conference on Artificial Intelligence and Law (ICAIL)}}. \bibinfo{pages}{159--168}.
\newblock


\bibitem[Zheng et~al\mbox{.}(2024)]%
        {zheng2024harnessing}
\bibfield{author}{\bibinfo{person}{Zhi Zheng}, \bibinfo{person}{Wenshuo Chao}, \bibinfo{person}{Zhaopeng Qiu}, \bibinfo{person}{Hengshu Zhu}, {and} \bibinfo{person}{Hui Xiong}.} \bibinfo{year}{2024}\natexlab{}.
\newblock \showarticletitle{Harnessing large language models for text-rich sequential recommendation}. In \bibinfo{booktitle}{\emph{Proceedings of the ACM on Web Conference 2024}}. \bibinfo{pages}{3207--3216}.
\newblock


\bibitem[Zhong et~al\mbox{.}(2024)]%
        {zhong2024logparser}
\bibfield{author}{\bibinfo{person}{Aoxiao Zhong}, \bibinfo{person}{Dengyao Mo}, \bibinfo{person}{Guiyang Liu}, \bibinfo{person}{Jinbu Liu}, \bibinfo{person}{Qingda Lu}, \bibinfo{person}{Qi Zhou}, \bibinfo{person}{Jiesheng Wu}, \bibinfo{person}{Quanzheng Li}, {and} \bibinfo{person}{Qingsong Wen}.} \bibinfo{year}{2024}\natexlab{}.
\newblock \showarticletitle{Logparser-llm: Advancing efficient log parsing with large language models}. In \bibinfo{booktitle}{\emph{Proceedings of the 30th ACM SIGKDD Conference on Knowledge Discovery and Data Mining}}. \bibinfo{pages}{4559--4570}.
\newblock


\bibitem[Zhu et~al\mbox{.}(2023)]%
        {zhu2023loghub}
\bibfield{author}{\bibinfo{person}{Jieming Zhu}, \bibinfo{person}{Shilin He}, \bibinfo{person}{Pinjia He}, \bibinfo{person}{Jinyang Liu}, {and} \bibinfo{person}{Michael~R Lyu}.} \bibinfo{year}{2023}\natexlab{}.
\newblock \showarticletitle{Loghub: A large collection of system log datasets for ai-driven log analytics}. In \bibinfo{booktitle}{\emph{2023 IEEE 34th International Symposium on Software Reliability Engineering (ISSRE)}}. IEEE, \bibinfo{pages}{355--366}.
\newblock


\end{thebibliography}
\balance

\end{document}